\documentclass[twocolumn,times]{aastex701}

\usepackage{xspace}
\usepackage{hyperref}

\newcommand{\msun}{$M_{\odot}$}

\newcommand{\mwd}{${M}_{\mathrm{WD}}$}
\newcommand{\teff}{${T}_{\mathrm{eff}}$}
\newcommand{\logg}{$\log{g}$}
\newcommand{\muhz}{$\mu$Hz}

\newcommand{\tar}{GD\,1212\xspace}

\shorttitle{Outburst of the DAV GD 1212}
\shortauthors{Hermes et al.}
\graphicspath{{./}{}}
\begin{document}

\title{Mode Instability and a Massive, Isolated Outburst in the Pulsating White Dwarf GD 1212}


\author[0000-0001-5941-2286]{J.~J.~Hermes}
\affiliation{Department of Astronomy \& Institute for Astrophysical Research, Boston University, 725 Commonwealth Ave., Boston, MA 02215, USA}
\email[show]{jjhermes@bu.edu}  

\author[0000-0002-0656-032X]{Keaton~J.~Bell}
\affiliation{Department of Physics, Queens College, City University of New York, Flushing, NY, 11367, USA}
\email{Keaton.Bell@qc.cuny.edu}

\author[0009-0008-2682-7088]{Andrew~H.~Dublin}
\affiliation{Department of Physics, Queens College, City University of New York, Flushing, NY, 11367, USA}
\email{adublin@gradcenter.cuny.edu}

\author[0000-0002-6748-1748]{M.~H.~Montgomery}
\affiliation{Department of Astronomy, University of Texas at Austin, 2515 Speedway, Austin, TX 78712, USA}
\email{mikemon@astro.as.utexas.edu}

\author[0000-0002-6536-6367]{Steven~D.~Kawaler}
\affiliation{Department of Physics and Astronomy, Iowa State University, Ames, IA 50011, USA}
\email{sdk@iastate.edu}

\author[0009-0002-6257-6649]{Ian~Clark}
\affiliation{Department of Physics and Astronomy, Iowa State University, Ames, IA 50011, USA}
\email{irclark@iastate.edu}

\author[0000-0002-0853-3464]{Zachary~P.~Vanderbosch}
\affiliation{Hobby-Eberly Telescope, 32 Mt. Locke Rd., McDonald Observatory, TX 79734, USA}
\email{zachary.vanderbosch@austin.utexas.edu}

\author[0000-0002-1086-8685]{Bart~H.~Dunlap}
\affiliation{Department of Astronomy, University of Texas at Austin, 2515 Speedway, Austin, TX 78712, USA}
\email{bhdunlap@utexas.edu}

\author[0000-0001-9873-0121]{P.-E.~Tremblay}
\affiliation{Department of Physics, University of Warwick, Coventry CV4~7AL, UK}
\email{P.Tremblay@warwick.ac.uk}

\author[0009-0006-2789-0893]{Paul~Chote}
\affiliation{Department of Physics, University of Warwick, Coventry CV4~7AL, UK}
\email{P.Chote@warwick.ac.uk}

\author[0000-0002-2761-3005]{Boris~T.~G\"{a}nsicke}
\affiliation{Department of Physics, University of Warwick, Coventry CV4~7AL, UK}
\email{boris.gaensicke@warwick.ac.uk}

\begin{abstract}

We analyze a large brightening event that lasted for roughly half a day in the pulsating hydrogen-atmosphere white dwarf GD\,1212 during K2 Campaign 12 of the extended Kepler mission. For the other 80 days of K2 observations, GD\,1212 exhibited a rich spectrum of long-period ($\sim$1100 s) pulsations that underwent rapid variations in frequency and amplitude but did not exhibit any additional outbursts. We refine previous attempts at mode identification and find a likely sequence of dipole and quadrupole splittings that reveal an overall rotation rate of roughly 17.0\,hr. The outburst at Day 61 is fully resolved by the 60-second-cadence K2 data, with the entire white dwarf becoming up to 17.5\% brighter overall (from an $\approx850$\,K increase in \teff), with pulsational variability during the outburst showing shorter periods and higher amplitudes. Outbursts are believed to be the result of nonlinear mode coupling via parametric instability, whereby energy stored in linearly excited parent modes is rapidly transferred to damped child modes that dissipate near the surface of the white dwarf. Additionally, we characterize a ``failed'' outburst that caused correlated pulsation frequency changes ($\approx5\,\mu$Hz increase) with a small ($\approx0.35$\%) corresponding brightness increase. GD\,1212 is now the eighth pulsating hydrogen-atmosphere (DAV) white dwarf to show outburst behavior, although it exhibited the largest outburst yet and has the longest inferred recurrence timescale. This high-signal-to-noise record tracing pulsations through both large and small temperature excursions in GD\,1212 provides unique insights into parametric resonance and nonlinear mode coupling in white dwarf pulsations.

\end{abstract}


\section{Introduction} \label{sec:intro}

Asteroseismology affords the opportunity to see below the photosphere and probe the internal constitution of pulsating stars. Space-based photometry from exoplanet missions like the Kepler space telescope has revolutionized this science, offering unprecedented light curve precision and duration that have yielded deep insights into the interiors of stars at many stages of their evolution  \citep{2021RvMP...93a5001A,2022afas.confE...1K}.

A critical step to matching the observed pulsation periods to theoretical models in order to perform asteroseismology is identifying the radial order ($n$), spherical degree ($\ell$), and azimuthal order ($m$) of the pulsation modes observed in the star \citep{2010aste.bookA}. For red giants and solar-like oscillators, the outer convection zone stochastically drives all modes over a frequency range spanning multiple consecutive overtones, allowing for pattern recognition using tools such as an \'{e}chelle diagram \citep{2013ARAA..51..353C}. This has led to major advances in constraining stellar interiors (e.g., \citealt{2012ApJ...748L..10M,2017ApJ...835..172L,2017ApJ...835..173S}).

For oscillating stars that are not stochastically driven, the set of observed modes is often incomplete, making the pattern recognition necessary for mode identification complicated. Pulsating white dwarf stars are a class of stars that are not stochastically driven \citep{1991MNRAS.251..673B,1999ApJ...511..904G,1999ApJ...519..783W}. Although white dwarfs were among the first types of star known to pulsate \citep{1968ApJ...153..151L}, progress on performing credible asteroseismic analysis on these stellar remnants has been hampered by this incomplete mode driving, despite extensive observational campaigns organized by the Whole Earth Telescope \citep[WET;][]{WET} in the 1990s (e.g., \citealt{1994ApJ...430..839W,1995ApJ...447..874K,1996A&A...314..182P}).

Pulsations in hydrogen-atmosphere white dwarfs (DAVs) are convectively driven non-radial gravity modes ($g$-modes), with longer-period modes preferentially excited as the surface convection zone deepens toward cooler effective temperatures \citep{2006ApJ...640..956M,2008PASP..120.1043F}. The $g$-modes are distributed about mean period spacings (the period between subsequent radial orders) of approximately 45\,s for $\ell=1$ modes and 25\,s for $\ell=2$ modes for canonical-mass (0.6\,\msun) white dwarfs, though this is highly dependent on effective temperature and hydrogen-layer mass \citep{2022A&A...663A.167A}. These period spacings cause a higher density of modes at low frequency. Each mode is also split into $2\ell+1$ components by rotation, typically at the level of $\sim$$5-10$\,$\mu$Hz \citep{2017ApJS..232...23H}.

The hottest DAVs exhibit the shortest-period pulsations and tend to show remarkable amplitude and phase stability; e.g., G117-B15A, has kept phase precision for more than 40 years of monitoring \citep{2021ApJ...906....7K}. This mode stability allows for the distinguishing of patterns in frequency spacing caused by rotation, often enabling reliable $\ell$ identification and thus mode identification; asteroseismology is thus possible for these hotter DAVs, although they tend to show fewer detected pulsations \citep{2014MNRAS.438.3086G,2016ApJS..223...10G}. However, as the surface convection zone that drives pulsations deepens, longer-period modes are driven that tend to appear less coherent, likely due to the outer turning point of modes longer than roughly 800\,s reaching to the base of the convection zone, which is constantly changing due to the pulsations \citep{2020ApJ...890...11M}. The coolest DAVs that show the most pulsation periods also show the most amplitude and phase modulation, complicating asteroseismic analysis.

Another feature of the coolest pulsating white dwarfs is flux outbursts, wherein the white dwarf gets up to tens of percent brighter in a matter of hours \citep{2015ApJ...809...14B,2015ApJ...810L...5H,2017ASPC..509..303B}. Outbursts are likely a rapid transfer of energy from driven parent modes to damped child modes via parametric resonance \citep{2001ApJ...546..469W,2018ApJ...863...82L}. Mode coupling via parametric resonance has been observed in the correlated frequency and amplitude modulation of modes in non-outbursting white dwarfs \citep{2016A&A...585A..22Z}. The conditions required to establish nonlinear limit cycles and produce outbursts likely favor cool white dwarfs with longer-period pulsations, where there is a higher density of modes available for three-mode couplings \citep{2001ApJ...546..469W}. It has yet to be shown if one parent mode or a cascade of multiple parent modes are responsible for a given outburst, and how the energy exchange is mediated. 

So far, all outbursts in pulsating white dwarfs have been detected by the Kepler space telescope, which had the sensitivity to detect a few percent brightening events for stars as faint as $K_p=19$\,mag. Outbursts in DAVs show no pattern or regularity to their recurrence time, and have been measured to occur as frequently as every 2.4 days \citep{2016ApJ...829...82B} and as infrequently as every 9.7 days \citep{2017ASPC..509..303B}.

\begin{figure*}[t!]
    \centering
    {\includegraphics[width=0.995\textwidth]{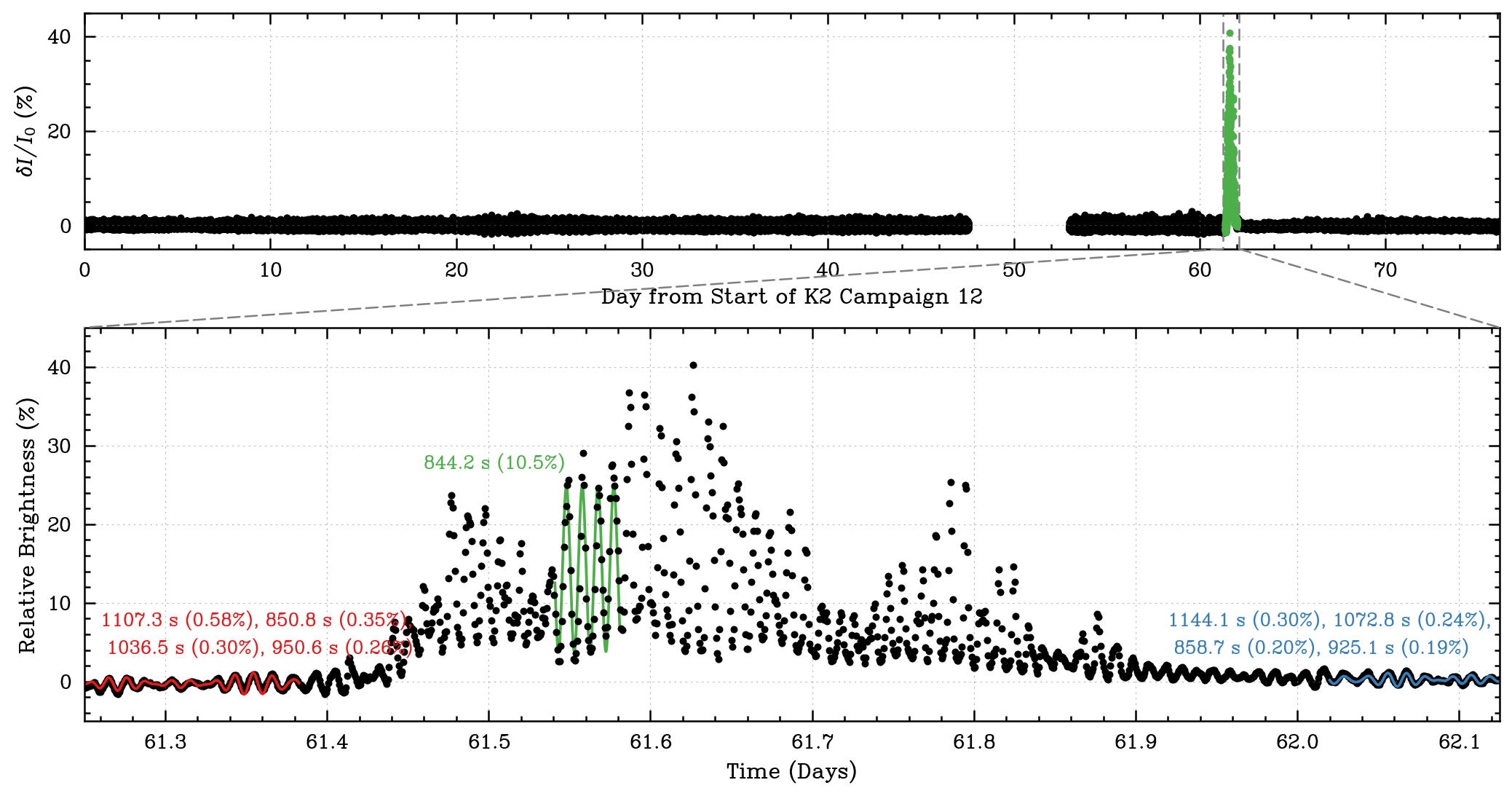}}
    \caption{The top panel shows the full 76.1-day K2 Campaign 12 light curve of \tar, with observations roughly every 60\,s. A large flux outburst is observed on Day 61; a zoom-in on the event is shown in the bottom panel. A sinusoidal model of the dominant pulsation periods in roughly 4-hr windows before, during, and after the outburst are labeled on the plot in red, green, and blue, respectively. The dominant pulsation signals in outburst have generally shorter periods and larger amplitudes, but return roughly to quiescent behavior in less than 1\,day.}%
    \label{fig:k2lightcurves}
\end{figure*}

\tar\ is a bright ($V=13.3$\,mag) white dwarf that has been known to pulsate for more than two decades \citep{2006AJ....132..831G}. It has spectroscopically determined atmospheric parameters of \teff\ $= 10{,}970\pm170$ K and \logg\ $= 8.03\pm0.05$ \citep{2011ApJ...743..138G}, and thus a mass of $0.62\pm0.03$ \msun\ using the 3D-model corrections of \citet{2013A&A...559A.104T}. This matches well the photometrically determined parameters of \teff\ $= 10{,}870\pm100$ K, \logg\ $= 8.01\pm0.02$, \mwd\ $=0.608\pm0.010$ \msun\ from \citet{2024MNRAS.527.8687O} that is constrained by the Gaia DR3 parallax. It is also
consistent with the other spectroscopic and photometric atmospheric constraints summarized by \citet{2017ApJ...851...60R} in the context of their asteroseismic analysis of \tar.

A $10{,}970$\,K spectroscopically derived effective temperature is cool enough to place \tar\ right among the known outbursting white dwarfs \citep{2017ApJS..232...23H}. However, no such outbursts were observed in more than 11 days of data collected in the so-called Campaign 0 initial engineering run of the K2 mission; the light curve and observations were described in detail in \citet{2014ApJ...789...85H}. We show here that \tar\ did not outburst for at least 61 days in subsequent observations in K2 Campaign 12 before undergoing a massive (up to 40\%) increase in flux that lasted for several hours.

In Section~\ref{sec:observations} we detail the K2 photometry and follow-up spectroscopy of \tar, and in Section~\ref{sec:outburst} we analyze the main flux outburst in K2 Campaign 12. Likely mode identification and pulsation variability are discussed in Section~\ref{sec:pulsationmodes}, including a possible ``failed outburst.'' We broaden our discussion and conclude in Section~\ref{sec:discuss}.

\section{Observations} \label{sec:observations}

We describe all new time-series photometric observations of \tar\ from the K2 mission, as well as follow-up spectroscopy to better constrain the atmospheric parameters.

\subsection{K2 Campaign 12 Photometry of GD 1212} \label{subsec:k2photometry}

The Kepler space telescope initially observed \tar\ for more than 11.5 days in early 2014 as part of an engineering test (so-called Campaign 0) of two-reaction-wheel controlled operations, and are described in detail in \citet{2014ApJ...789...85H}. \tar\ was re-observed in K2 Campaign 12 (from 15 December 2015 to 4 March 2016), comprising the primary dataset analyzed here. We included \tar\ in our target list for short-cadence (58.85-second exposures) of known or candidate pulsating white dwarfs (GO12040). Roughly 47.5 days into the observations the spacecraft entered a safe mode, likely due to a flight software reset, creating a nearly 5.5-day gap in data collection.

We have extracted the K2 photometry following the procedures described in \citet{2017ApJS..232...23H}. In short, we downloaded the short-cadence Target Pixel File for the target (EPIC 246074853, $K_p = 13.3$\,mag) from the Mikulski Archive for Space Telescopes and extracted the photometry with a 17-pixel fixed aperture using the \texttt{PyKE} software suite \citep{2012ascl.soft08004S}, using the \texttt{KEPSFF} tool to detrend K2 roll-induced systematics \citep{2014PASP..126..948V}. The quality of the photometry is exquisite. We clipped 869 single-point outliers by visual inspection as being significantly deviant from the typical 0.2\% point-to-point scatter of the dataset. This yields an overall duty cycle of 91.9\% over 76.1 days (the duty cycle is 99.0\% outside of the large gap induced by the safe-mode event). All times and phases reported in this manuscript are relative to the $t_0$ of observations BJD$_{\rm TDB}=$ 2457738.3674321. We also analyze the \texttt{K2SFF} high-level science product based on the 30-min-cadence data of \tar\ during Campaign 12 \citep{2014PASP..126..948V} in Section~\ref{sec:outburst} and Section~\ref{subsec:failedoutburst}.

Our full light curve is shown in Figure~\ref{fig:k2lightcurves}, including a zoom-in around the large flux outburst on Day 61. As with another bright outbursting DAV, PG\,1149+057 \citep{2015ApJ...810L...5H}, pulsations are clearly visible before, during, and after the outburst in the bottom panel of Figure~\ref{fig:k2lightcurves}. We compute periodograms in 4-hr windows before, during, and after the outburst and plot the best-fit significant signals as a sum of sinusoids shown in Figure~\ref{fig:k2lightcurves}. That the pulsation amplitudes appear larger in outburst demonstrates the outburst is on the white dwarf and not the brightening of another background star within the extraction aperture.

\begin{figure*}[t!]
    \centering
    {\includegraphics[width=0.9\textwidth]{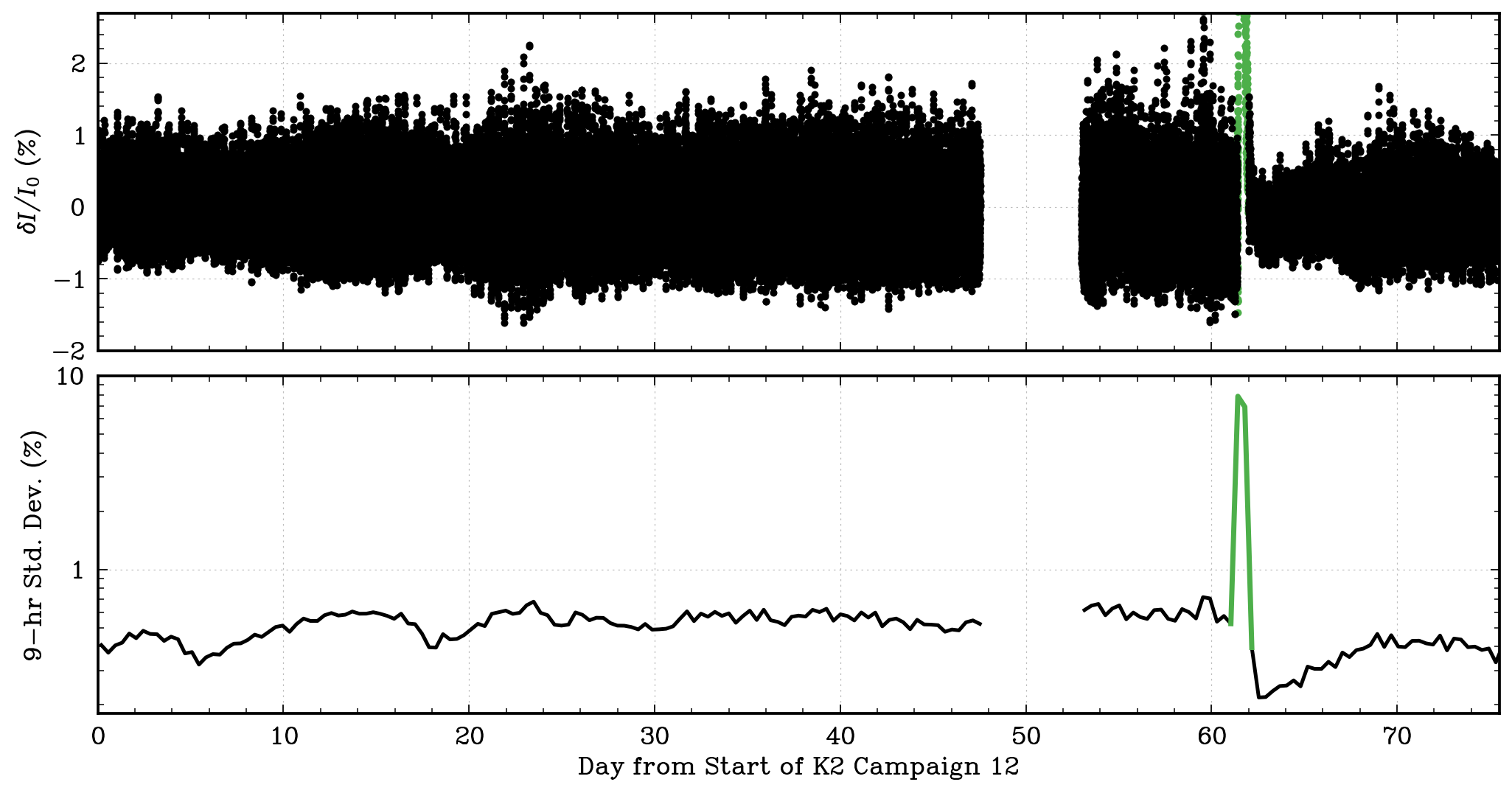}}
    \caption{The top panel shows the full short-cadence K2 dataset of \tar. As can be seen in more detail in the bottom panel of Figure~\ref{fig:k2lightcurves}, the pulsations dominate the out-of-outburst light curve, with roughly $2-3$\% peak-to-peak amplitudes. The bottom panel shows the standard deviation of the short-cadence flux computed in consecutive 9-hour bins (with a logarithmic y-axis). The overall pulsation amplitudes vary significantly over the dataset, and are significantly diminished after the outburst at Day 61, slowly growing back for the week that follows.}%
    \label{fig:zoomdeplete}
\end{figure*}

\subsection{Low-Resolution SOAR Spectroscopy} \label{subsec:soarspectroscopy}

There is extensive optical spectroscopy of \tar\ in the literature. In fact, aside from G29$-$38 (e.g., \citealt{2000MNRAS.314..209V}), \tar\ might have the most time-series spectroscopy collected of any DAV, as it was observed over four consecutive nights in October 2006 with the Boller \& Chivens spectrograph on the 2.3-meter Bok telescope at Kitt Peak \citep{Desgranges2008}. The four nights of Bok observations were an attempt at mode identification of the pulsations using line-profile changes of 431 separate 100-second spectra of \tar, but were inconclusive. Still, the coadded spectra provided excellent constraints on the atmospheric parameters of the star; the data were analyzed most recently by \citet{2011ApJ...743..138G}.

New spectroscopy of \tar\ from the Southern Astrophysical Research (SOAR) telescope in Chile was presented in \citet{2017ApJS..232...23H}, which showed similar atmospheric parameters: \teff\ $= 10{,}980\pm140$ K and \logg\ $= 8.00\pm0.04$. However, these observations (collected on 2016 Jul 14) were only five 30-second exposures spanning less than 200 seconds, much shorter than the pulsation periods, which we should optimally average over.

We obtained additional spectroscopy on SOAR using the Goodman spectrograph \citep{2004SPIE.5492..331C} with the 930-line grating; we obtained 25$\times$60-second spectra on 2017 November 9 and 12$\times$120-second spectra on 2017 November 26. Both sequences cover more than a full pulsation period. Data were obtained and reduced in an identical manner as \citet{2017ApJS..232...23H}, and fit using the same models of \citet{2011ApJ...730..128T} using ML2/$\alpha=0.8$.

Fits to the first night of observations find \teff\ $= 11{,}060\pm130$ K and \logg\ $= 8.02\pm0.03$ and thus mass of $0.61\pm0.02$ \msun. Similarly, fits to the second new night of data find \teff\ $= 10{,}970\pm130$ K and \logg\ $= 8.01\pm0.03$ and thus mass of $0.61\pm0.02$ \msun. Both fits include the 3D-model corrections of \citet{2013A&A...559A.104T}.

The new spectroscopy is entirely consistent with previous observations, so we use the weighted mean of all four values to refine the baseline atmospheric parameters of \tar: \teff\ $= 11{,}000\pm130$ K and \logg\ $= 8.012\pm0.030$ and thus mass of $0.614\pm0.020$ \msun.

\section{A Massive Outburst at Day 61} \label{sec:outburst}

The most dramatic feature in the light curve is the unmistakable flux outburst 61.4 days into the K2 Campaign 12 observations, highlighted in green in the top panel of Figure~\ref{fig:k2lightcurves}. \tar\ returns to varying about the median relative flux of the full dataset by Day 62.0 (the pulsations can still be seen in the blue fit in Figure~\ref{fig:k2lightcurves}), but the overall amplitude of the variability at that point has significantly diminished. This is shown in Figure~\ref{fig:zoomdeplete}, which reveals that the range of flux variability has significantly decreased after the outburst.

We quantify this decrease by computing the standard deviation of the short-cadence (58.85-s exposures) data in consecutive 9-hour bins, shown in the bottom panel of Figure~\ref{fig:zoomdeplete}. For the week before the outburst, this standard deviation held constant near roughly 0.6\% amplitude. For the 2\,d after the outburst (Days $62-64$) this value dropped below 0.25\%, a factor of 2.5 decrease.

The standard deviation within running 9-hr intervals reveals that the overall pulsation amplitudes of \tar\ go through significant amplitude changes from week to week. This standard deviation reaches as low as 0.35\% at Day 6 and 0.4\% at Day 18, to as high as $>$8.0\% during the outburst itself. The overall pulsation amplitudes appear to be near a maximum for more than a week before the massive outburst at Day 61.

Since the pulsational variability increased in outburst, we analyze the 30-min-cadence light curve extracted by \texttt{K2SFF} to measure the maximum global flux increase; the 30-min cadence washes out even the longest-period pulsations and yields an estimate of the total surface flux increase in outburst of roughly 17.5\% higher than the typical quiescent flux during the rest of K2 Campaign 12. To estimate the energy released in outburst and the corresponding increase in effective temperature, we utilize the synthetic photometric magnitudes \citep{2006AJ....132.1221H} derived from model spectra of DA white dwarf atmospheres from \citet{2011ApJ...730..128T}.\footnote{\url{https://www.astro.umontreal.ca/~bergeron/CoolingModels/}} Kepler bandpass magnitudes ($K_p$) are not precomputed in those model grids, so we estimate them based on model $g$ and $r$ magnitudes using the relation for blue targets ($g - r \leq  0.8$) provided by the Kepler mission:\footnote{\url{https://nexsci.caltech.edu/workshop/2012/keplergo/CalibrationZeropoint.shtml}; accurate to $\pm0.2$\,mag.} $K_p = 0.2 g + 0.8 r$.

Taking the spectroscopic \teff\ of $11{,}000$\,K as the quiescent value, we can interpolate from the models a \teff\ estimate for a given relative flux increase in the Kepler band. At the peak overall flux increase of 17.5\%, the white dwarf photosphere reaches $T_{\rm eff} \approx 11{,}850$\,K. Such an increase in effective temperature would still keep the white dwarf within the bounds of the empirical DAV instability strip \citep{2015ApJ...809..148T}, but would significantly thin the outer convection zone; the increased amplitude of pulsations around 844.2\,s seen in Figure~\ref{fig:k2lightcurves} is likely the result of decreased attenuation from the convection zone \citep{1991MNRAS.251..673B}. 

Assuming no change in radius of the white dwarf, we use the Stefan-Boltzmann law to calculate the increase in luminosity as the effective temperature changes. Integrating the excess luminosity above the quiescent value of $8.24\times10^{30}\,{\rm erg\,s}^{-1}$ throughout the outburst yields an estimated total energy released in the outburst of $7.5\times10^{34}$\,erg. This is the most energetic outburst yet published, by a factor of four \citep{2016ApJ...829...82B}.

\section{The Pulsation Spectrum of GD 1212} \label{sec:pulsationmodes}

Determining a reliable census of observed oscillation modes for white dwarfs with pulsations that are unstable in amplitude has been a long-standing problem. For example, the DBV GD\,358 has been monitored in detail by multiple WET runs and shows some modes entirely absent in some observing runs that are present in others \citep{1994ApJ...430..839W,2009ApJ...693..564P,2010ApJ...716...84M}. This is also the case for the well-studied DAV G29-38 \citep{1998ApJ...495..424K,2023MNRAS.526.2846U}.

\begin{figure*}[t!]
    \centering
    {\includegraphics[width=0.995\textwidth]{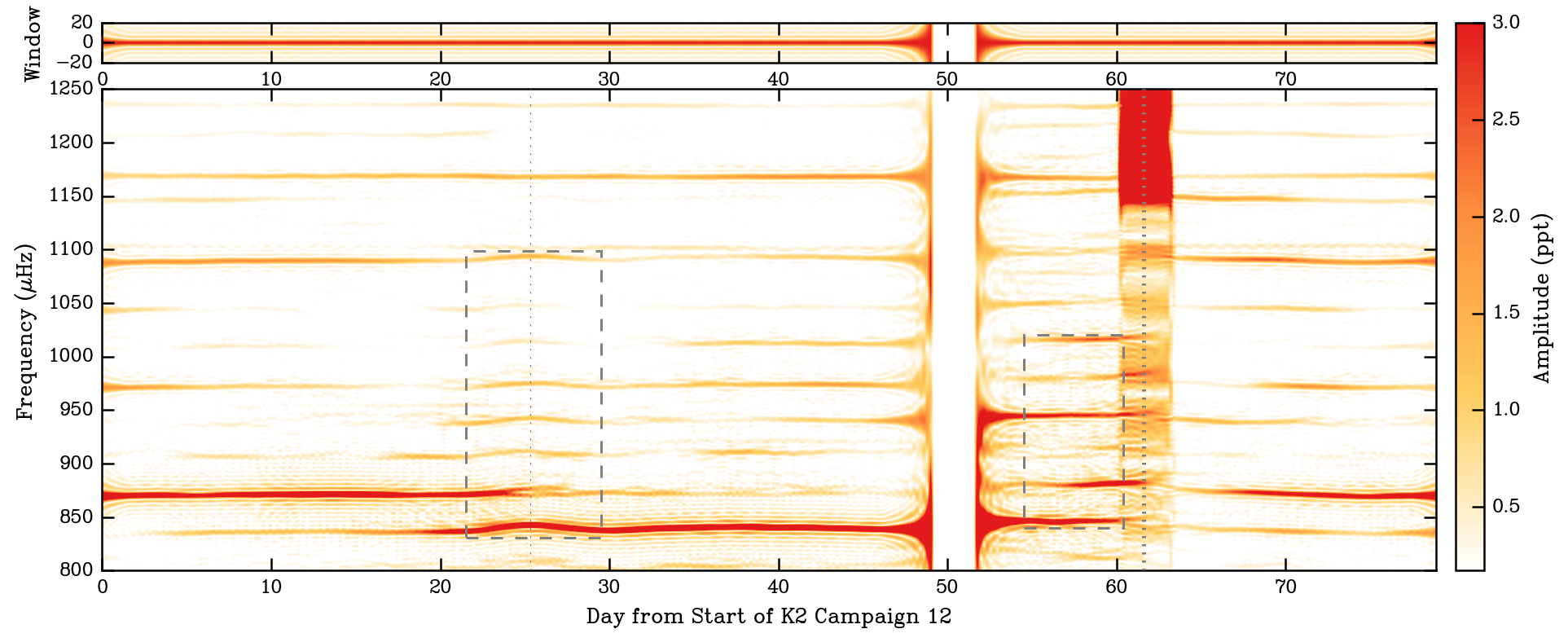}}
    \caption{Running periodogram of the full K2 Campaign 12 light curve of \tar, focused on the independent pulsation modes present from $800-1250$\,\muhz. We use a 3-day sliding window, which smears events to that resolution. The window function in the top panel is at the same y-axis scale and shows we can resolve frequencies within 4\,\muhz. Pulsation amplitudes during the outburst at Day 61 move dramatically; dominant modes are near 870\,\muhz\ (1150\,s) before the outburst, but shift to higher frequencies nearer 1185\,\muhz\ (844\,s) during the outburst. A region of a possible failed outburst around Day 25 is detailed further in Section~\ref{subsec:failedoutburst}. }
    \label{fig:runningFT}
\end{figure*}

\tar\ exhibits major changes in its pulsation spectrum, as discussed in the first analysis of the periodogram of the first 9 days of K2 observations obtained in K2 Campaign 0 \citep{2014ApJ...789...85H}. The pulsation mode variability of \tar\ can be seen in the running periodogram shown in Figure~\ref{fig:runningFT}. The apparent frequency broadening near the large data gap around Day~50 is caused by the sliding-window analysis rather than intrinsic pulsation variability: as the 3-day window approaches the gap, fewer data points contribute to each periodogram, producing the same broadening in the window function shown in the top panel of Figure~\ref{fig:runningFT}.

\begin{figure*}[t!]
    \centering
    {\includegraphics[width=0.995\textwidth]{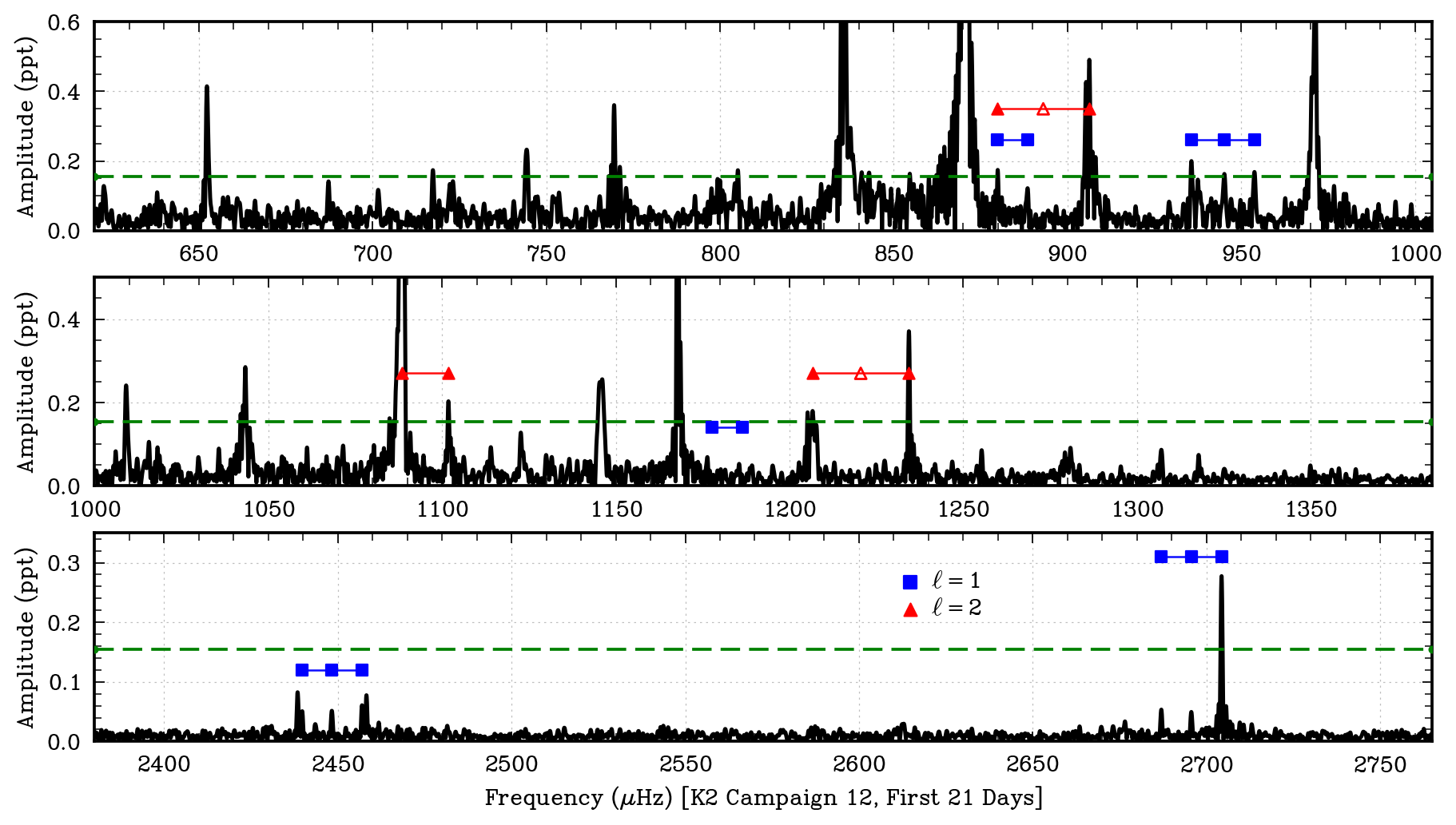}}
    \caption{Static periodogram of the first 21 days of the K2 Campaign 12 light curve of \tar\ in three frequency ranges, chosen as it is the longest interval exhibiting the greatest apparent stability in the K2 data. Colored markers indicate candidate multiplet identifications from our period list, with blue squares for likely $\ell=1$ modes (see Table~\ref{tab:id1}) and red triangles for $\ell=2$ modes. Lines connect the expected $m$ components using mean rotational splittings of $\delta\nu_{\ell=1}=8.66$\,\muhz\ and $\delta\nu_{\ell=2}=13.6$\,\muhz\ based on an overall rotation rate of roughly 17.0\,hr (see Section~\ref{subsec:rotation}). Filled symbols are observed, while open symbols show adjacent azimuthal orders that are not observed in this dataset. All peaks above the green dashed line for the significance threshold are detailed in Appendix~\ref{app:periods}. }
    \label{fig:staticFT}
\end{figure*}

The running periodogram in Figure~\ref{fig:runningFT} was generated by computing 400 periodograms with a sliding 3-day window of data, and covers the range of the independent pulsation frequencies from $800-1250$\,\muhz\ (which coincidentally corresponds to periods from $1250-800$\,s). Our resolution limit on both the x- and y-axes in Figure~\ref{fig:runningFT} is set by the 3-day sliding window. In the first half of the dataset, mode amplitudes can be seen growing and shrinking on timescales of days. The mode instability makes generating a census of identified pulsation periods extremely difficult. The lack of phase coherence of most modes creates broad peaks in a static periodogram over the first 21 days (the longest interval exhibiting the greatest apparent stability in the K2 data), shown in Figure~\ref{fig:staticFT}. We consider all peaks above 5.4$\times$ the mean value in the static periodogram between $500-3000$\,\muhz\ as significant \citep{2015MNRAS.448L..16B}, and include all identified peaks above this value (dashed green line in Figure~\ref{fig:staticFT}) in Appendix~\ref{app:periods}.

\begin{figure}[t!]
    \centering
    {\includegraphics[width=0.995\columnwidth]{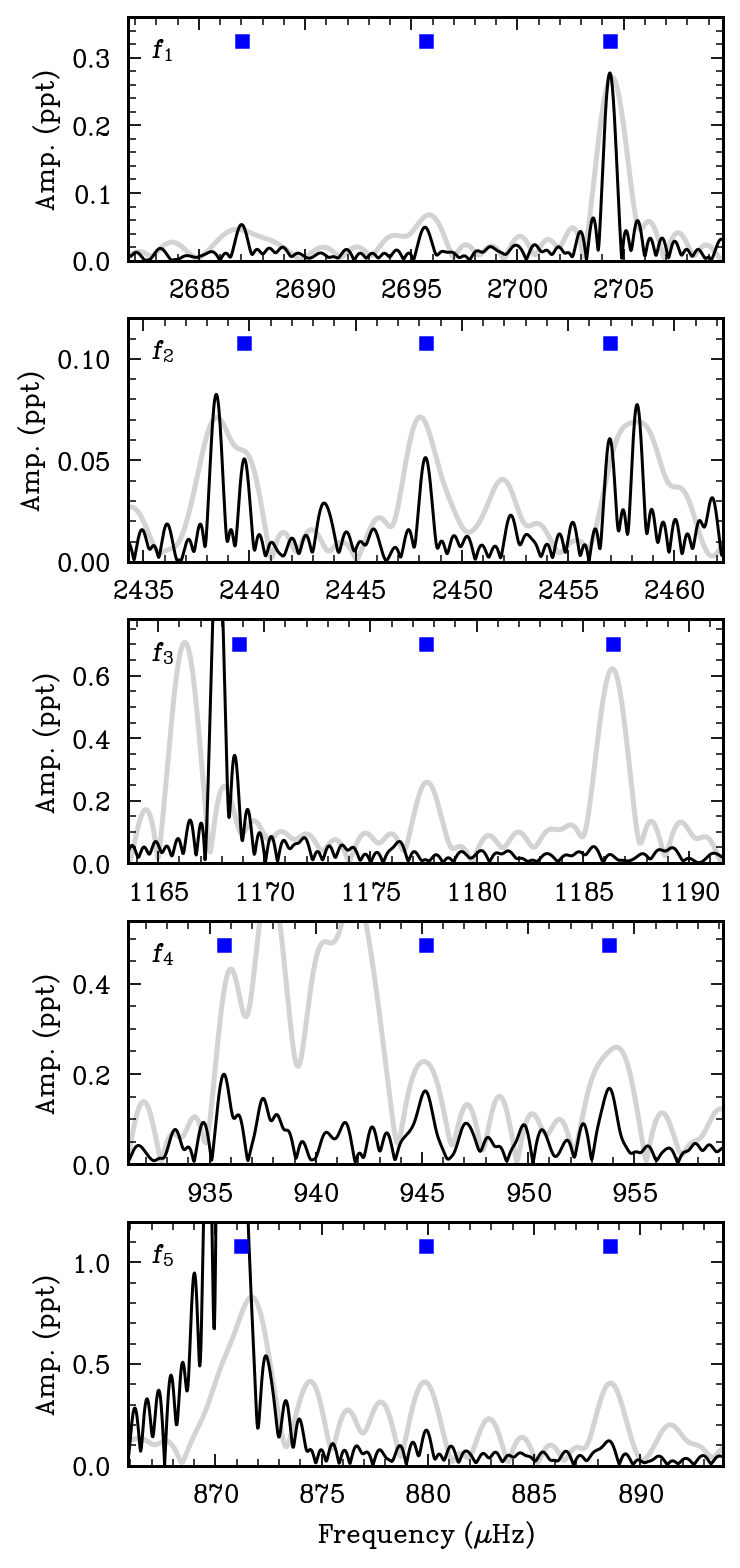}}
    \caption{Amplitude spectra of \tar\ showing five frequency windows centered on likely $\ell=1$ modes, ordered from highest frequency (top) to lowest frequency (bottom) matching the order in Table~\ref{tab:id1}. In each panel, light gray curves show power spectra from selected 9-day segments of the K2 Campaign 0 or Campaign 12 light curves, while the black curve highlights the first 21 days of K2 Campaign 12.}
    \label{fig:multiplets}
\end{figure}

Figure~\ref{fig:staticFT} shows that there are several modes that appear consistent with $\ell=1$ multiplets, marked with dark blue squares. Additionally, there are many modes that are separated by integer numbers of the 13.6\,\muhz\ frequency spacing expected for $\ell=2$ modes in a white dwarf rotating at 17.0\,hr. For the five $\ell=1$ multiplets that appear symmetric and readily identifiable, we propose the mode identification and mean splitting in Table~\ref{tab:id1}.

Just as in the analysis of the Campaign 0 data described by \citet{2014ApJ...789...85H}, we observe independent pulsations that dominate in the frequency range between roughly $600-1400$\,\muhz, as well as in the range of $2400-2740$\,\muhz\ (corresponding to periods from $1670-715$\,s and $415-365$\,s, respectively). There are also nonlinear combination frequencies (e.g., $2f_1$ or $f_1+f_2$) at frequencies ranging from $1600-2200$\,\muhz, as well as difference frequencies (e.g., $f_3-f_2$) at very low frequencies, $<$$300$\,\muhz. We focus here on the independent frequencies that are natural standing modes in the star.

\subsection{Rotational Multiplets Suggest 17-hr Rotation} \label{subsec:rotation}

While mode identification was not attempted by \citet{2014ApJ...789...85H}, further analysis presented in \citet{2017ApJS..232...23H} found patterns in the Campaign 0 data that appeared to show $\ell=1$ splittings averaging around 21.3\,\muhz\ and $\ell=2$ splittings of roughly 33.7\,\muhz, both roughly consistent with a 7.0-hr rotation period. We find here that with more data from Campaign 12 this mode identification is incorrect, an unlucky coincidence of similar patterns in the periodogram. In this period range, the asymptotic period spacing maps into quasi-regular frequency spacings of tens of \muhz, making overtone structure in frequency space comparable to rotational multiplet splitting structure, the likely source of the previous incorrect mode identification. 

The K2 Campaign 12 data reveal that there are two significant high-frequency multiplets, centered at 2695.7\,\muhz\ (370.96\,s) and 2448.3\,\muhz\ (408.45\,s), both of which appear to be $\ell=1$ triplets with roughly even splittings of 8.66\,\muhz\ and are shown in the top two panels of Figure~\ref{fig:multiplets}. These higher-frequency modes are relatively easier to identify since the rotational splittings are much smaller than the overtone spacing. Assuming a Ledoux constant for $\ell=1$ modes of $C_{n,1}=0.47$ (see discussion in \citealt{2017ApJS..232...23H}\footnote{We use $n$ for radial order but many white dwarf studies use $k$.}), a $\delta \nu_{\ell=1} = 8.66$\,\muhz\ implies an overall rotation rate for \tar\ of 17.0\,hr, and a predicted $\delta \nu_{\ell=2} = 13.6$\,\muhz, assuming $C_{n,2}=0.167$. This rotation period follows from the first-order Ledoux splitting relation, $\delta\nu_{n\ell m}=m(1-C_{n\ell})\Omega/2\pi$, evaluated for adjacent $m$ components.

\begin{figure*}[t!]
    \centering
    {\includegraphics[width=0.995\textwidth]{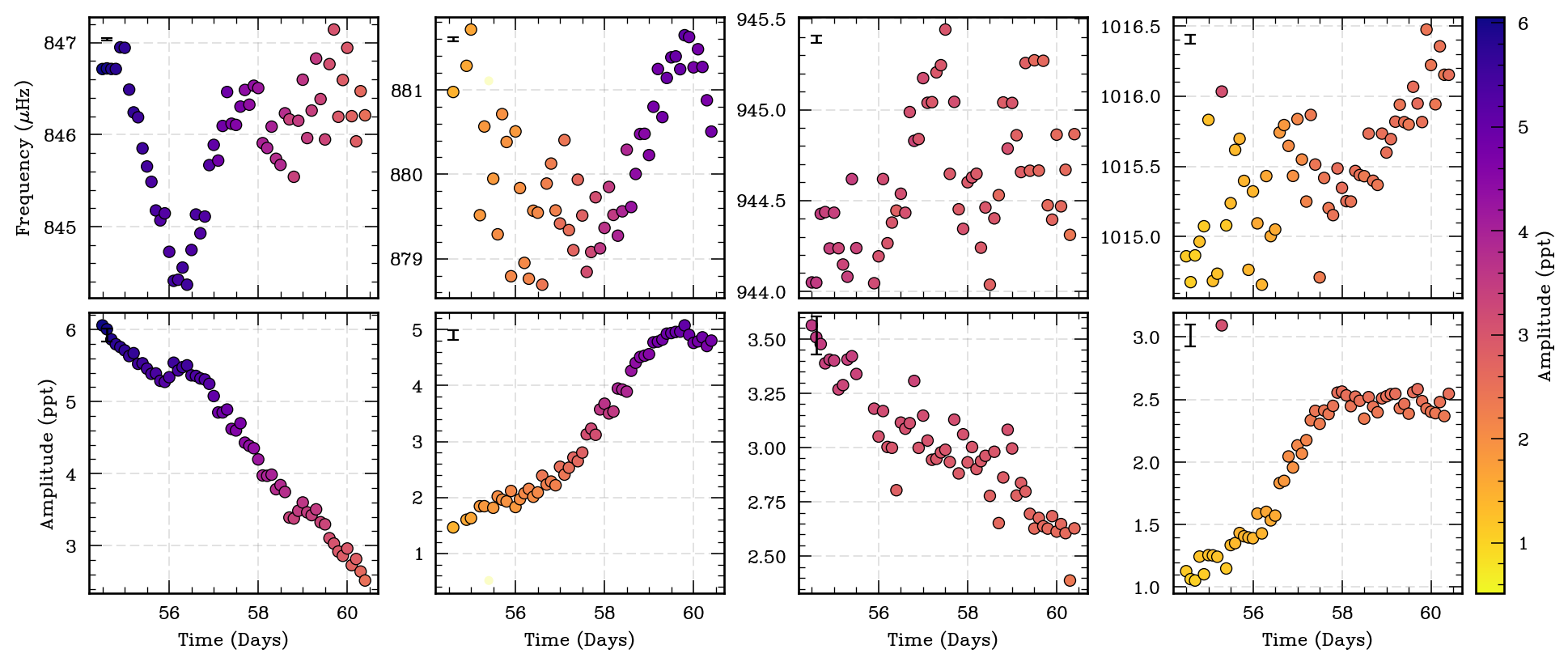}}
    \caption{Time evolution of four of the highest-amplitude independent pulsation frequencies (top row) and their corresponding amplitudes (bottom row) fit to subsets of the data in a 2-day sliding window across Days 53.5–61.4, immediately before the large outburst at Day 61. The first and last modes centered at roughly 846 and 1015 \muhz\ are not present in Figure~\ref{fig:staticFT} for the first 21 days of K2 Campaign 12; the middle modes centered at roughly 880 and 945 \muhz\ are likely $\ell=1,m=0$ modes identified as $f_{5}$ and $f_{4}$ in Table~\ref{tab:id1}. There is strong, correlated frequency and amplitude variability within many modes in the star, likely due to strong mode coupling and parametric resonances \citep{2016A&A...585A..22Z}. Mean uncertainties for each point are indicated in the upper left of each subplot. }
    \label{fig:runpreoutburst}
\end{figure*}

\begin{table}
\caption{Likely $m=0$ central components of $\ell=1$ pulsation multiplets present in first 21 days of K2 Campaign 12 for \tar\ detailed in Figure~\ref{fig:multiplets}.}
\label{tab:id1}
\centering
\begin{tabular*}{\columnwidth}{@{\extracolsep{\fill}}lccc}
\hline\hline
ID & Frequency & Period & $\delta\nu$ \\
   & ($\mu$Hz) & (s)    &  ($\mu$Hz) \\
\hline
$f_{1}$ & $2695.68\pm0.14$ & 370.96 & $8.67\pm0.12$  \\
$f_{2}$ & $2448.30\pm0.13$ & 408.45 & $8.65\pm0.14$ \\
$f_{3}$ & $1177.62\pm0.15$ & 849.2  & $8.78\pm0.16$  \\
$f_{4}$ & $945.16\pm0.04$ & 1058.0 & $8.68\pm0.06$  \\
$f_{5}$ & $879.80\pm0.05$ & 1136.7 & $8.66\pm0.08$  \\
\hline
\end{tabular*}
\end{table}

The most stable modes, as in most cool DAVs, are those at the highest frequencies (see \citealt{2020ApJ...890...11M}). Strangely, the $f_2$ mode at 408.45\,s (second panel in Figure~\ref{fig:multiplets}) has two extra peaks that are separated from the central component by an additional 1.24\,\muhz\ that we refer to as  ``outrigger'' peaks. We do not have a physical explanation for these ``outrigger'' peaks in $f_2$ but they appear outside of both the prograde and retrograde $m=\pm1$ components.

We define the frequencies of our five $\ell=1$ multiplets using the first 21 days of Campaign 12, but also show in gray in Figure~\ref{fig:multiplets} data from shorter 9-day segments of various K2 light curves. While some of the 21-day signals are not significant compared to our 0.155\,ppt significance threshold (see Appendix~\ref{app:periods}), each shorter 9-day segment has a significant peak near a component marked with a blue square in Figure~\ref{fig:multiplets}. For example, in K2 Campaign 0, there are significant frequencies at $1177.62\pm0.15$\,\muhz\ and $1186.40\pm0.06$\,\muhz\ \citep{2014ApJ...789...85H}\footnote{These were defined as the $\ell=2$ component $f_{5b}$ and the $\ell=1$ component $f_{3b}$, respectively, in the incorrect mode identification of \citealt{2017ApJS..232...23H}.}.

Additionally, there are many components in Figure~\ref{fig:staticFT} that fit the pattern of spacings expected for components of $\ell=2$ multiplets. We observe significant peaks in the first 21 days of data in Campaign 12 at 1088.58\,\muhz\ and 1101.84\,\muhz, corresponding to $\delta \nu_{\ell=2} = 13.26\pm0.04$\,\muhz. Similarly, we likely observe a skipped azimuthal order in modes at 1206.60\,\muhz\ and 1234.38\,\muhz, corresponding to $\delta \nu_{\ell=2} = 13.89\pm0.10$\,\muhz. Thus, \tar\ is likely pulsating in a mix of $\ell=1$ and $\ell=2$ multiplets.

It is possible there are also regions with very closely overlapping modes of different multiplets (e.g., \citealt{2012MNRAS.420.1462R,2022A&A...663A.167A}). For example, the central component of the $\ell=1$ mode $f_5$ at 879.80\,\muhz\ is separated by $\delta \nu_{\ell=2} = 13.25\pm0.06$\,\muhz\ from a significant peak at 906.29\,\muhz, and so may also be very close to another component of an $\ell=2$ multiplet. These close frequency overlaps may work to strongly enhance resonance and nonlinear mode coupling in the star, since limit cycles happen most frequently when the frequency mismatch is minimized, which happens more often as the mode density increases \citep{2001ApJ...546..469W}. As we show in the next sections, all modes in \tar\ with periods longer than 800\,s show significant amplitude and frequency changes, further complicating mode identification.

\subsection{Pulsation Incoherence in GD 1212} \label{subsec:instability}

As visualized in the running periodogram in Figure~\ref{fig:runningFT}, the highest-amplitude pulsations in \tar\ are highly unstable in amplitude from week to week during the K2 Campaign 12 observations. Less visible is the extent to which the pulsation frequencies actually evolve.

We attempt to better visualize both the frequency and amplitude variability of select modes in Figure~\ref{fig:runpreoutburst}, where we show the evolution of four of the highest-amplitude pulsations over nearly 6 days before the massive outburst at Day 61.4. Our sliding 2-day window smears out features to that scale. For every step in time, we perform a nonlinear least-squares fit for the frequency, amplitude, and phase of a sinusoid,  using the frequency of the highest peak in windows running from 830--850\,\muhz, 864--884\,\muhz, 930--950\,\muhz, and 1000--1020\,\muhz.

Figure~\ref{fig:runpreoutburst} reveals that both the amplitudes and frequencies experience significant variability on extremely short timescales: amplitudes can increase or decrease by a factor of 3 in less than a week, and frequencies can wander by more than 2.5\,\muhz\ in less than a week. The mode density for these long-period modes is high, but not infinite; adiabatic models by \citet{2012MNRAS.420.1462R} suggest there are likely to be two $\ell=1$ modes and three $\ell=2$ modes centered between $800-850$\,\muhz\ (so roughly nine $\ell=1$ multiplet components and 15 $\ell=2$ components). The $\ell=1$ multiplet spacing of 8.66\,\muhz\ suggests it is unlikely for adjacent pulsations to be alternately excited and de-excited to explain these frequency shifts, as the multiplets are too widely separated.

Instead, the mode variability is likely best explained by mode coupling and nonlinear resonance, which has been observed to operate in pulsating white dwarfs (e.g., \citealt{2013ApJ...766...42H,2016A&A...585A..22Z}). However, unlike the more regular amplitude/frequency correlations observed by \citet{2016A&A...585A..22Z} in a DBV, \tar\ shows larger and less coherent changes. This may indicate a denser network of nonlinear mode couplings, or a more strongly nonlinear regime in which many coupled modes contribute simultaneously to the observed variability. Stochastic driving and damping associated with convection will also contribute to the observed loss of phase coherence, broadening Fourier peaks on timescales comparable to the intrinsic mode growth and decay times \citep{2015ApJ...809...14B,2020ApJ...890...11M}. However, the steady evolution and large frequency changes in Figure~\ref{fig:runpreoutburst} likely require a stronger mechanism, since growth rates are typically of order days for modes $>$800\,s \citep{2001ApJ...546..469W}. We likely see the limit cycle behavior of this parametric resonance manifest as an outburst, but even before the outburst there appears to be significant steady-state instability at most times.

The large outburst begins at Day 61.4, so with our 2-day sliding window, data after Day 60.4 are affected by the outburst. It is difficult to see a direct correlation of any one pulsation mode and the outburst onset. The mode with the largest amplitude decrease is that nearest 846\,\muhz, though this is a mode we have not been able to identify given its rapid frequency changes.

The likely $\ell=1,m=0$ mode at 879.8\,\muhz\ appears to vary more than $\pm1$\,\muhz\ and strongly increase in amplitude for the week before the large outburst, while the likely $\ell=1,m=0$ mode at 945.2\,\muhz\ is steadily decreasing in amplitude. Essentially all high-amplitude modes with periods longer than 800\,s are unstable in frequency and amplitude when observed for more than a week, though these changes differ from mode to mode.

\subsection{Correlated Frequency Shifts in a ``Failed'' Outburst} \label{subsec:failedoutburst}

\begin{figure}[t!]
    \centering
{\includegraphics[width=0.92\columnwidth]{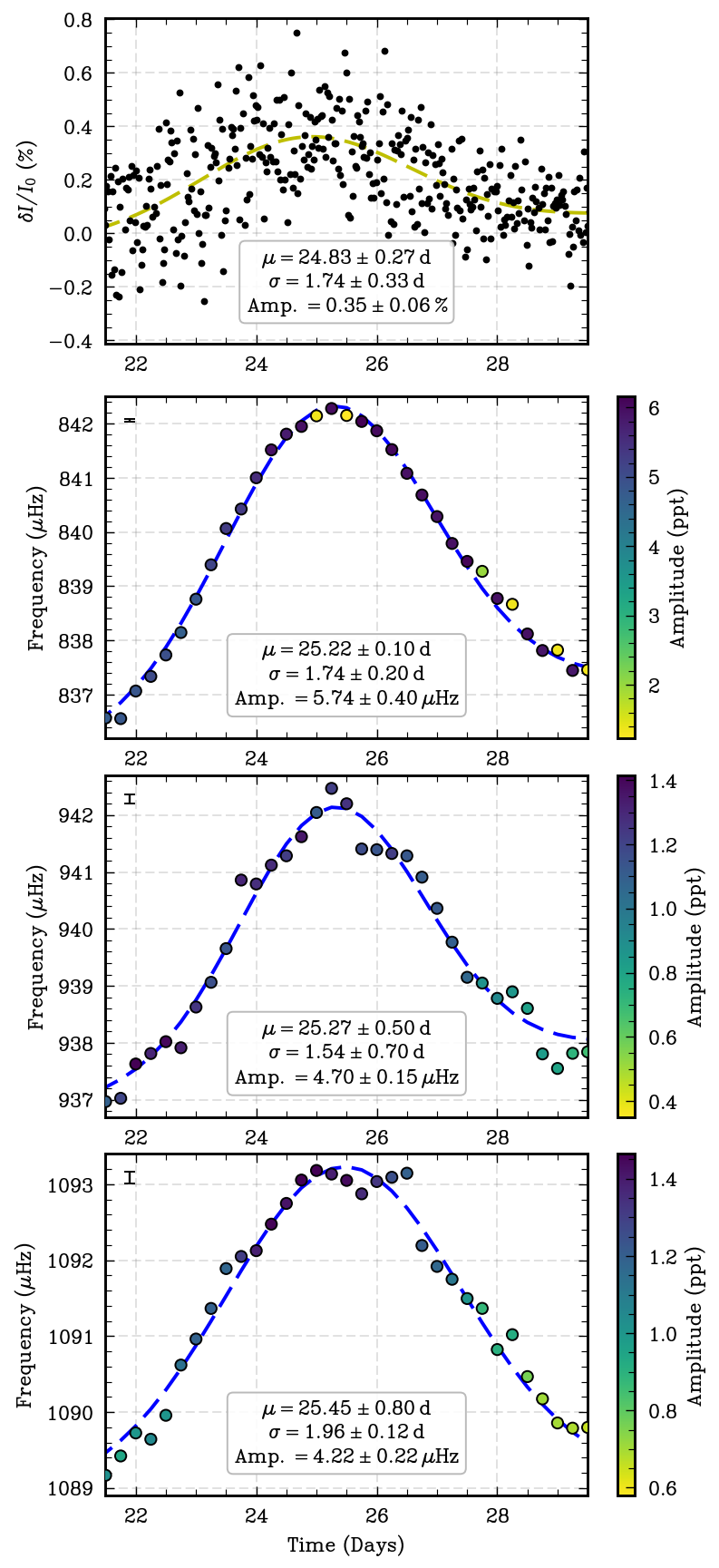}}
    \caption{Time evolution of three independent pulsation modes measured across a 3-day sliding window (see first gray box in Figure~\ref{fig:runningFT}). All three modes shift to higher frequencies by $\approx5\,\mu\mathrm{Hz}$, with lower-frequency modes shifting by a larger amount. The dashed blue lines show joint line-plus-Gaussian fits; the parameters of the Gaussian component are reported in each panel. Mean uncertainties for each point are indicated in the upper left of each subplot. These correlated frequency shifts provide evidence for a common underlying physical process modulating the pulsation frequencies during a ``failed'' outburst near Day~25; the top panel shows the associated small-scale brightening event in the 30-min-cadence light curve that averages over the pulsations.}  %
    \label{fig:threefailed}
\end{figure}

Aside from the dramatic change in behavior around Day 61 caused by the large outburst (Section~\ref{sec:outburst}), there is also a much smaller increase in flux around Day 25. Although it is not obvious in Figure~\ref{fig:k2lightcurves}, Figure~\ref{fig:threefailed} shows a long-timescale flux increase in \tar\ of up to 0.35\% in the 30-min-cadence \texttt{K2SFF} light curve, which smooths over the short-period pulsations. The running periodogram shown in Figure~\ref{fig:runningFT} reveals a concurrent shift of many of the lowest-frequency modes in the star to higher frequencies. 

The lower panels of Figure~\ref{fig:threefailed} trace the shifts in frequency of the three highest-amplitude modes fitted within a 3-day sliding window, which occur simultaneously with the 0.35\% overall increase in flux in the \textit{Kepler} bandpass. Fitting a Gaussian to the \texttt{K2SFF} light curve between Days $18-32$ yields a midpoint at Day $24.83\pm0.09$ and a peak amplitude of 0.35\%. The Gaussian width corresponds to a duration (FWHM) of approximately 4.1\,d. Because this is much smaller than the impulsive flux increase of 17.5\% of the large outburst and we can track the gradual, multi-day evolution of the individual pulsation modes, we consider that this may be a gentler version of the same phenomenon, and we discuss this event as a  ``failed outburst.''

The three signals shown to modulate similarly in frequency are distinct pulsation modes; since we only expect $\ell=1,2$ modes to be visible in DAVs (e.g., \citealt{2000MNRAS.314..220C}), any features in the periodogram separated by more than 60\,\muhz\ cannot arise from the same multiplet and must be from independent pulsation modes in the star. We have quantified these correlated changes by jointly fitting linear and Gaussian components to each frequency evolution diagram in Figure~\ref{fig:threefailed}, reporting the central component of the Gaussian ($\mu$), the standard deviation (Gaussian width) of the transient in time ($\sigma$), and the frequency-shift amplitude of each event. The amplitude of the shift for all modes exceeds 4\,\muhz, and the lowest-frequency modes appear to shift by a greater amount. All modes reach their frequency maximum at the same time, Day 25.3, and are statistically consistent with an event that has a duration of $4.45\pm0.24$\,d (FWHM). The apparent duration may be inflated by the width of the 3-day window used to fit the frequencies, although using smaller sliding windows (down to 1.0\,d) only slightly decreases the FWHM (to 4.0\,d).

One possible phenomenological interpretation for both the brightness increase and the correlated shift of these mode frequencies is a temporary increase in \teff\, potentially caused by the sudden damping of pulsational energy into the outer layers of the star. This could be a similar mechanism to that suspected to trigger massive outbursts like that observed at Day 61 \citep{2018ApJ...863...82L}. In this case, the disruption may not be sufficient to cause a runaway cascade of mode damping to trigger a massive outburst, leading only to a ``failed outburst.'' Using the same approach as in Section~\ref{sec:outburst} to estimate the 17.5\% overall flux increase in the \textit{Kepler} bandpass to correspond to an $\approx850$\,K increase in effective temperature, we find that a 0.35\% flux increase should correspond to a temperature increase of only $\approx16$\,K. Integrating the excess flux yields an energy estimate of $1.7\times10^{34}$\,erg, assuming this is caused entirely by a global increase in effective temperature. While spread over a longer period of time, this would be nearly one-quarter the energy contained in the outburst at Day 61 and comparable in energy to other outbursting DAVs (e.g., \citealt{2015ApJ...810L...5H}).

The depth of the outer convection zone is extremely sensitive to temperature and rapidly becomes shallower in response to even a small temperature change. The correlated frequency shift in multiple independent modes could potentially be explained by the mode cavity changing due to an increase in effective temperature. Resonant frequencies of all $\ell=1$ modes with periods $\gtrsim800$\,s ($\lesssim1250$\,\muhz) that have their outer boundary conditions set by the base of the convection zone \citep{2020ApJ...890...11M} would increase in tandem while the excess energy is radiated away.

\begin{figure}[t!]
    \centering
    {\includegraphics[width=0.995\columnwidth]{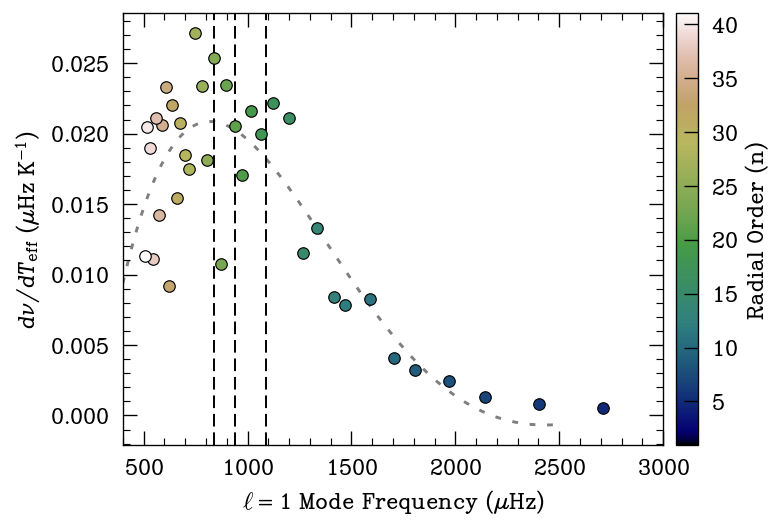}}
    \caption{We use the white dwarf modeling code WDEC to compute the expected rate of frequency change with effective temperature for different $\ell=1$ modes as a function of frequency in a representative model for \tar. The models attempt to capture mode cavity changes due to a changing convection zone depth in response to superficial heating (see Appendix~\ref{app:wdec}). The three modes traced through the ``failed outburst'' in Figure~\ref{fig:threefailed} are marked with vertical dashed lines, and a fifth-order polynomial captures the overall trend. The calculated rates of frequency change would require an effective temperature increase of $\sim250$\,K to match the observed shifts of $\approx5$\,\muhz, which is inconsistent with the 0.35\% brightness increase ($\approx16$\,K) observed by K2 in Figure~\ref{fig:threefailed}. 
}%
    \label{fig:df_dTeff_l=1}
\end{figure}

We use the White Dwarf Evolution Code (WDEC, \citealt{WDEC}) to explore how frequencies might change as the convection zone thins due to an increase in effective temperature, with more details provided in Appendix~\ref{app:wdec}. In short, we modify the depth of the convection zone by varying the $\alpha$ parameter for the ML2/$\alpha$ mixing length theory for convection to mimic a change in effective temperature due to superficial heating. We measure rates of frequency change for models with parameters close to those from spectroscopy (0.62\,\msun, $10{,}970$\,K) assuming a canonically thick hydrogen-layer mass ($M_{\mathrm{H}}/M_\star = 10^{-4}$). 
Figure~\ref{fig:df_dTeff_l=1} shows the numerical rate of change of individual mode frequencies with effective temperature, $d\nu/dT_{\mathrm{eff}}$, as a function of mode frequency. The frequencies of the three modes traced in Figure~\ref{fig:threefailed} are marked with vertical lines. 

A low-order polynomial summarizes the overall trend that lower frequency $\ell=1$ modes change most rapidly in the range of observed frequencies, while high-frequency modes that are not bounded by the base of the convection zone are unaffected. This trend matches the ratios of fitted frequency shifts reported in Figure~\ref{fig:threefailed}, though individual model modes show significant scatter about the overall trend. Appendix~\ref{app:wdec} shows that corresponding shifts of $\ell=2$ mode frequencies are expected to be nearly twice as large and generally show a different overall trend for lower-frequency modes.

\begin{figure*}[t!]
    \centering
    {\includegraphics[width=0.95\textwidth]{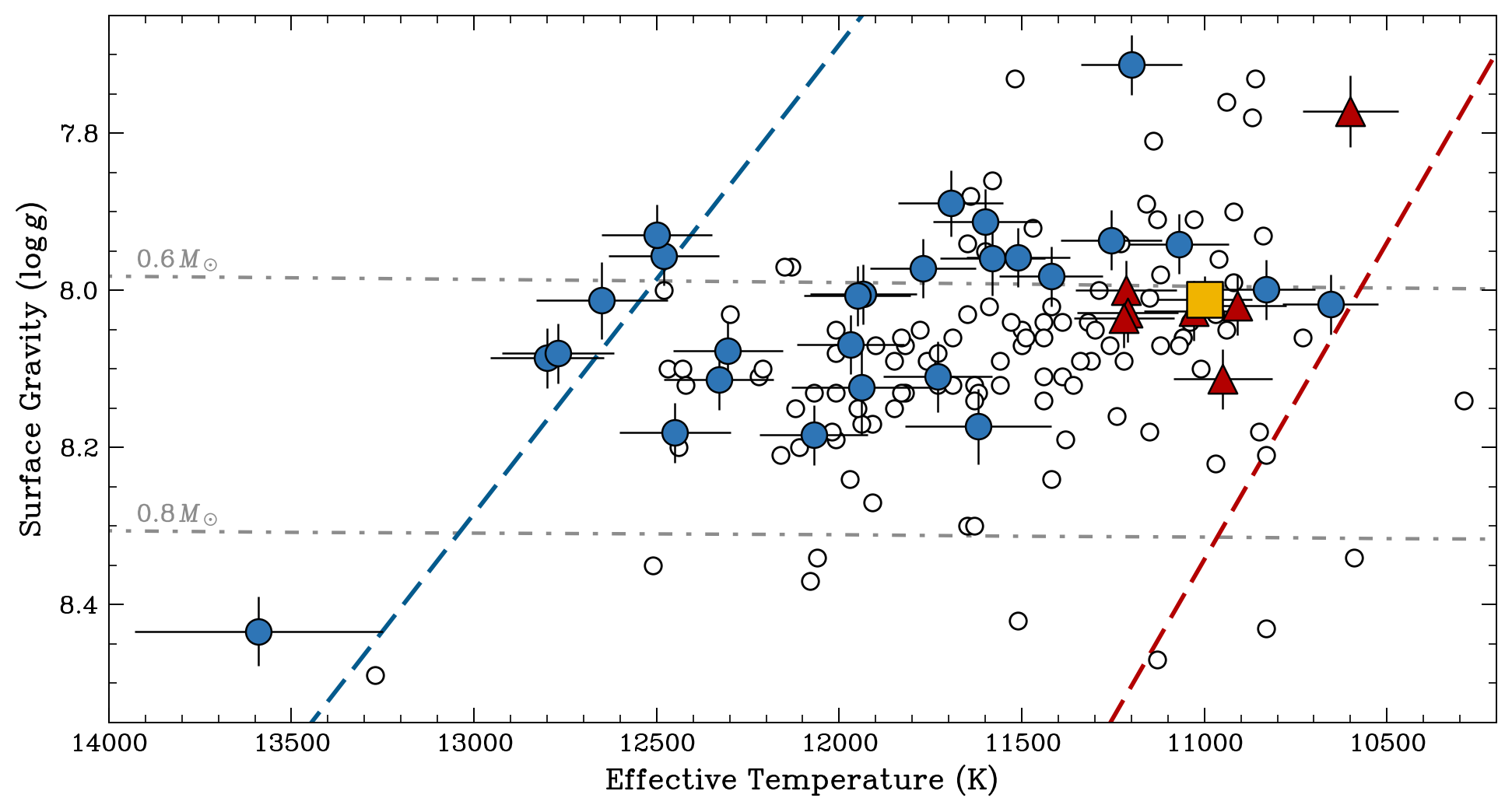}}
    \caption{Following \citet{2017ApJS..232...23H}, we show the strip of pulsating white dwarfs, with \tar\ marked as an orange square. Pulsators observed by Kepler/K2 are marked as blue circles, and those with detected outbursts shown as red triangles; ground-based detections of pulsations are shown with open circles.}%
    \label{fig:istrip}
\end{figure*}

To match the magnitude of the $\approx5$\,\muhz\ frequency shifts observed for modes from $835-1095$\,\muhz\ (Figure~\ref{fig:threefailed}), the model values of $d\nu/dT_{\rm eff}\approx0.02\,\mu$Hz\,\teff$^{-1}$ (Figure~\ref{fig:df_dTeff_l=1}) would require a $\sim250$\,K overall increase in effective temperature. Such a temperature increase would be expected to produce an $\approx5.3\%$ flux increase in the Kepler bandpass. That significantly over-predicts the observed overall 0.35\% flux increase.

Thus, there remains a quantitative discrepancy, by about a factor of 15, between the heating inferred from the overall flux increase and that implied by the rates of frequency change with effective temperature that we calculate with our WDEC models as described in Appendix~\ref{app:wdec}. This discrepancy may imply that our approach to modeling transient surface heating effects with a white dwarf envelope model like WDEC is too simplified to be quantitatively accurate.

While a plausible picture emerges in which a modest (0.35\%) increase in flux corresponds to an overall heating event that shrinks the convection zone and causes correlated increases in pulsation frequencies during this ``failed outburst,'' such large frequency shifts we observe ($>$4\,\muhz) are not expected for a $<20$\,K heating event. This discrepancy may imply that an entirely different mechanism is responsible.

\section{Discussion \& Conclusions} \label{sec:discuss}

The K2 Campaign 12 observations of \tar\ reveal one of the most dramatic flux outbursts yet detected in a pulsating hydrogen-atmosphere white dwarf. The event at Day 61 produces a $\simeq$$17.5\%$ increase in the baseline stellar flux, while short-cadence K2 data demonstrate that pulsations persist throughout the outburst. \tar\ is thus the eighth known DAV to exhibit outburst behavior, but it is unusual in showing only a single clear event over more than 80 days of high-duty-cycle Kepler monitoring, implying one of the longest outburst recurrence times in the class. \tar\ is the brightest known outbursting DAV, allowing pulsations to be traced even in  short subsets of the light curve.

The atmospheric parameters of \tar\ (\teff\ $= 11{,}000\pm130$ K and \logg\ $= 8.012\pm0.030$ and thus mass of $0.614\pm0.020$ \msun) put it squarely in the middle of the other known outbursting DAVs, as shown in Figure~\ref{fig:istrip}. Though it has the longest recurrence timescale seen so far in the outbursting DAVs, it does not appear to be the hottest of the class, suggesting that mode density and other selection effects are important in determining how frequently a DAV undergoes an outburst event.

The dominant phenomenology of the Day~61 outburst matches previous outburst observations in DAVs: a rapid brightening on hours-long timescales, a return to quiescence in less than 0.5\,d, and a contemporaneous shift of pulsation power toward shorter periods (higher frequencies) with enhanced amplitudes. In all cases, outbursts appear qualitatively consistent with being produced by nonlinear mode coupling and parametric resonance \citep{2001ApJ...546..469W,2018ApJ...863...82L}. In this framework, energy stored in driven parent $g$-modes is rapidly transferred to damped child modes once a parametric instability threshold is crossed. Dissipation of the child mode energy near the surface leads to a transient increase in emergent flux. The outburst is followed by a pronounced depletion of overall variability, with pulsation amplitudes lower by a factor of at least $\simeq 2.5$ immediately after the outburst, which then slowly grow back.

Changes to the outer boundary conditions can modify the apparent pulsation periods; as the effective temperature increases, the convection zone shrinks, shifting modes to higher frequency, preferentially driving higher-frequency modes, and increasing mode visibility. The $\simeq17.5\%$ overall flux increase at Day 61 is estimated to come from an $\approx850$\,K temperature increase, during which the dominant pulsations are observed to have higher frequencies and higher amplitudes. A ``failed'' outburst near Day 25 may be a miniature version of the same phenomenon, where a much smaller global flux increase (0.35\%) is tracked by substantial ($\approx5$\,\muhz) correlated frequency shifts of low-frequency modes. Our attempts to model frequency changes due to superficial heating show qualitative agreement with the observed trends, but quantitatively implies a temperature increase of $\sim250$\,K that should cause a $\approx5.3\%$ increase in flux in the Kepler bandpass, which we do not see in the data. These events may be examples of the same phenomenon observed at two energy scales, providing valuable clues toward a better theoretical description of outburst energy exchange.

The observed pulsation frequencies and amplitudes are constantly changing in \tar, making it extremely difficult to provide a census of all observed pulsations, let alone complete mode identification. Still, we see patterns in the frequencies that reveal at least five modes that are well-described as $\ell=1$ multiplets, implying an overall mean rotation period of 17.0\,hr. 

Future progress will benefit from theoretical modeling of nonlinear correlations in the observed frequency/amplitude evolution of modes. Observational constraints from extended monitoring by missions such as the Transiting Exoplanet Survey Satellite (TESS) afford the ability to better constrain the outburst recurrence timescale in \tar\ (Clark et al., in prep.). Ultimately, the outbursting DAVs not only hold immense promise for studying the interplay of convection, mode coupling, and nonlinear amplitude saturation in white dwarf pulsations, but also the poorly understood cessation of pulsations at the red edge of the DAV instability strip.

\section*{Acknowledgments}

The authors acknowledge A.~Bischoff-Kim for guidance about the use of WDEC for this work, as well as the six-pack of Firemans \#4 bestowed upon Joshua T.\ Fuchs, who correctly predicted outbursts in \tar\ based on his PhD thesis work. Support for this work was in part provided by NASA K2 Cycle 4 grant NNX17AE92G, K2 Cycle 6 grant 80NSSC19K0162, as well as TESS Cycle 7 grants 80NSSC25K7902 and 80NSSC25K0122. K.J.B.\ and A.H.D.\ are supported by the National Science Foundation under grant No.\ AST-2406917. Funding for the Kepler mission is provided by the NASA Science Mission Directorate. M.H.M. and B.H.D. acknowledge support from the Wootton Center for Astrophysical Plasma Properties, a U.S. Department of Energy NNSA Stewardship Science Academic Alliance Center of Excellence supported under award numbers DE-NA0003843 and DE-NA0004149, from the United States Department of Energy under grant DE-SC0010623. M.H.M. acknowledges support from the NASA ADAP program under grant 80NSSC20K0455.

\vspace{5mm}
\facilities{Kepler (K2), SOAR (Goodman)}
\software{astropy \citep{2013A&A...558A..33A,2018AJ....156..123A}, Matplotlib \citep{Hunter:2007}, NumPy \citep{harris2020array}, pandas \citep{pandas2020}, SciPy \citep{2020SciPy-NMeth}, Pyriod \citep{2022ascl.soft07007B}, Period04 \citep{2005CoAst.146...53L}, WDEC \citep{WDEC}, MESA \citep{2011ApJS..192....3P}, GYRE \citep{2013MNRAS.435.3406T} }

\bibliography{GD1212_cleaned}{}

@ARTICLE{2000MNRAS.314..220C,
       author = {{Clemens}, J.~C. and {van Kerkwijk}, M.~H. and {Wu}, Y.},
        title = "{Mode identification from time-resolved spectroscopy of the pulsating white dwarf G29-38}",
      journal = {\mnras},
         year = 2000,
        month = may,
       volume = {314},
       number = {2},
        pages = {220-228},
          doi = {10.1046/j.1365-8711.2000.03307.x},
archivePrefix = {arXiv},
       eprint = {astro-ph/9911216},
 primaryClass = {astro-ph},
       adsurl = {https://ui.adsabs.harvard.edu/abs/2000MNRAS.314..220C}
}

@ARTICLE{2013ApJ...766...42H,
       author = {{Hermes}, J.~J. and {Montgomery}, M.~H. and {Mullally}, Fergal and {Winget}, D.~E. and {Bischoff-Kim}, A.},
        title = "{A New Timescale for Period Change in the Pulsating DA White Dwarf WD 0111+0018}",
      journal = {\apj},
         year = 2013,
        month = mar,
       volume = {766},
       number = {1},
          eid = {42},
        pages = {42},
          doi = {10.1088/0004-637X/766/1/42},
archivePrefix = {arXiv},
       eprint = {1302.1875},
 primaryClass = {astro-ph.SR},
       adsurl = {https://ui.adsabs.harvard.edu/abs/2013ApJ...766...42H}
}

@ARTICLE{2000MNRAS.314..209V,
       author = {{van Kerkwijk}, M.~H. and {Clemens}, J.~C. and {Wu}, Y.},
        title = "{Surface motion in the pulsating DA white dwarf G29-38}",
      journal = {\mnras},
         year = 2000,
        month = may,
       volume = {314},
       number = {2},
        pages = {209-219},
          doi = {10.1046/j.1365-8711.2000.02931.x},
archivePrefix = {arXiv},
       eprint = {astro-ph/9911215},
 primaryClass = {astro-ph},
       adsurl = {https://ui.adsabs.harvard.edu/abs/2000MNRAS.314..209V}
}

@INPROCEEDINGS{2004SPIE.5492..331C,
   author = {{Clemens}, J.~C. and {Crain}, J.~A. and {Anderson}, R.},
    title = "{The Goodman spectrograph}",
booktitle = {Ground-based Instrumentation for Astronomy},
     year = 2004,
   series = {\procspie},
   volume = 5492,
   editor = {{Moorwood}, A.~F.~M. and {Iye}, M.},
    month = sep,
    pages = {331-340},
      doi = {10.1117/12.550069},
   adsurl = {http://adsabs.harvard.edu/abs/2004SPIE.5492..331C}
}

@ARTICLE{2011ApJ...730..128T,
   author = {{Tremblay}, P.-E. and {Bergeron}, P. and {Gianninas}, A.},
    title = "{An Improved Spectroscopic Analysis of DA White Dwarfs from the Sloan Digital Sky Survey Data Release 4}",
  journal = {\apj},
archivePrefix = "arXiv",
   eprint = {1102.0056},
 primaryClass = "astro-ph.SR",
     year = 2011,
    month = apr,
   volume = 730,
      eid = {128},
    pages = {128},
      doi = {10.1088/0004-637X/730/2/128},
   adsurl = {http://adsabs.harvard.edu/abs/2011ApJ...730..128T}
}

@ARTICLE{2015ApJ...809..148T,
       author = {{Tremblay}, P.-E. and {Gianninas}, A. and {Kilic}, M. and {Ludwig}, H.-G. and {Steffen}, M. and {Freytag}, B. and {Hermes}, J.~J.},
        title = "{3D Model Atmospheres for Extremely Low-mass White Dwarfs}",
      journal = {\apj},
         year = 2015,
        month = aug,
       volume = {809},
       number = {2},
          eid = {148},
        pages = {148},
          doi = {10.1088/0004-637X/809/2/148},
archivePrefix = {arXiv},
       eprint = {1507.01927},
 primaryClass = {astro-ph.SR},
       adsurl = {https://ui.adsabs.harvard.edu/abs/2015ApJ...809..148T}
}

@mastersthesis{Desgranges2008,
  author       = {Guy Desgranges},
  title        = {Astérosismologie de l’étoile naine blanche variable GD 1212},
  school       = {Universit\'e de Montr\'eal},
  year         = {2008},
  type         = {M\'emoire de ma\^{i}trise},
  url          = {https://umontreal.scholaris.ca/items/8b5a76dd-8d1c-4b3b-8b48-a542a456315a},
}

@ARTICLE{2010ApJ...716...84M,
       author = {{Montgomery}, M.~H. and {Provencal}, J.~L. and {Kanaan}, A. and {Mukadam}, Anjum S. and {Thompson}, S.~E. and {Dalessio}, J. and {Shipman}, H.~L. and {Winget}, D.~E. and {Kepler}, S.~O. and {Koester}, D.},
        title = "{Evidence for Temperature Change and Oblique Pulsation from Light Curve Fits of the Pulsating White Dwarf GD 358}",
      journal = {\apj},
         year = 2010,
        month = jun,
       volume = {716},
       number = {1},
        pages = {84-96},
          doi = {10.1088/0004-637X/716/1/84},
archivePrefix = {arXiv},
       eprint = {1004.3579},
 primaryClass = {astro-ph.SR},
       adsurl = {https://ui.adsabs.harvard.edu/abs/2010ApJ...716...84M}
}

@ARTICLE{2009ApJ...693..564P,
       author = {{Provencal}, J.~L. and {Montgomery}, M.~H. and {Kanaan}, A. and {Shipman}, H.~L. and {Childers}, D. and {Baran}, A. and {Kepler}, S.~O. and {Reed}, M. and {Zhou}, A. and {Eggen}, J. and {Watson}, T.~K. and {Winget}, D.~E. and {Thompson}, S.~E. and {Riaz}, B. and {Nitta}, A. and {Kleinman}, S.~J. and {Crowe}, R. and {Slivkoff}, J. and {Sherard}, P. and {Purves}, N. and {Binder}, P. and {Knight}, R. and {Kim}, S. -L. and {Chen}, Wen-Ping and {Yang}, M. and {Lin}, H.~C. and {Lin}, C.~C. and {Chen}, C.~W. and {Jiang}, X.~J. and {Sergeev}, A.~V. and {Mkrtichian}, D. and {Andreev}, M. and {Janulis}, R. and {Siwak}, M. and {Zola}, S. and {Koziel}, D. and {Stachowski}, G. and {Paparo}, M. and {Bognar}, Zs. and {Handler}, G. and {Lorenz}, D. and {Steininger}, B. and {Beck}, P. and {Nagel}, T. and {Kusterer}, D. and {Hoffman}, A. and {Reiff}, E. and {Kowalski}, R. and {Vauclair}, G. and {Charpinet}, S. and {Chevreton}, M. and {Solheim}, J.~E. and {Pakstiene}, E. and {Fraga}, L. and {Dalessio}, J.},
        title = "{2006 Whole Earth Telescope Observations of GD358: A New Look at the Prototype DBV}",
      journal = {\apj},
         year = 2009,
        month = mar,
       volume = {693},
       number = {1},
        pages = {564-585},
          doi = {10.1088/0004-637X/693/1/564},
archivePrefix = {arXiv},
       eprint = {0811.0768},
 primaryClass = {astro-ph},
       adsurl = {https://ui.adsabs.harvard.edu/abs/2009ApJ...693..564P}
}

@ARTICLE{2017ApJ...851...60R,
       author = {{Romero}, Alejandra D. and {C{\'o}rsico}, A.~H. and {Castanheira}, B.~G. and {De Ger{\'o}nimo}, F.~C. and {Kepler}, S.~O. and {Koester}, D. and {Kawka}, A. and {Althaus}, L.~G. and {Hermes}, J.~J. and {Bonato}, C. and {Gianninas}, A.},
        title = "{Probing the Structure of Kepler ZZ Ceti Stars with Full Evolutionary Models-based Asteroseismology}",
      journal = {\apj},
         year = 2017,
        month = dec,
       volume = {851},
       number = {1},
          eid = {60},
        pages = {60},
          doi = {10.3847/1538-4357/aa9899},
archivePrefix = {arXiv},
       eprint = {1711.01338},
 primaryClass = {astro-ph.SR},
       adsurl = {https://ui.adsabs.harvard.edu/abs/2017ApJ...851...60R}
}

@ARTICLE{2006ApJ...640..956M,
       author = {{Mukadam}, Anjum S. and {Montgomery}, M.~H. and {Winget}, D.~E. and {Kepler}, S.~O. and {Clemens}, J.~C.},
        title = "{Ensemble Characteristics of the ZZ Ceti Stars}",
      journal = {\apj},
         year = 2006,
        month = apr,
       volume = {640},
       number = {2},
        pages = {956-965},
          doi = {10.1086/500289},
archivePrefix = {arXiv},
       eprint = {astro-ph/0507425},
 primaryClass = {astro-ph},
       adsurl = {https://ui.adsabs.harvard.edu/abs/2006ApJ...640..956M}
}

@ARTICLE{2015MNRAS.448L..16B,
       author = {{Baran}, A.~S. and {Koen}, C. and {Pokrzywka}, B.},
        title = "{A detection threshold in the amplitude spectra calculated from Kepler data obtained during K2 mission.}",
      journal = {\mnras},
         year = 2015,
        month = mar,
       volume = {448},
        pages = {L16-L19},
          doi = {10.1093/mnrasl/slu194},
archivePrefix = {arXiv},
       eprint = {1412.2753},
 primaryClass = {astro-ph.SR},
       adsurl = {https://ui.adsabs.harvard.edu/abs/2015MNRAS.448L..16B}
}

@ARTICLE{1998ApJ...495..424K,
       author = {{Kleinman}, S.~J. and {Nather}, R.~E. and {Winget}, D.~E. and {Clemens}, J.~C. and {Bradley}, P.~A. and {Kanaan}, A. and {Provencal}, J.~L. and {Claver}, C.~F. and {Watson}, T.~K. and {Yanagida}, K. and {Nitta}, A. and {Dixson}, J.~S. and {Wood}, M.~A. and {Grauer}, A.~D. and {Hine}, B.~P. and {Fontaine}, G. and {Liebert}, James and {Sullivan}, D.~J. and {Wickramasinghe}, D.~T. and {Achilleos}, N. and {Marar}, T.~M.~K. and {Seetha}, S. and {Ashoka}, B.~N. and {Mei{\v{s}}tas}, E. and {Leibowitz}, E.~M. and {Moskalik}, P. and {Krzesi{\'n}ski}, J. and {Solheim}, J. -E. and {Bruvold}, A. and {O'Donoghue}, D. and {Kurtz}, D.~W. and {Warner}, B. and {Martinez}, Peter and {Vauclair}, G. and {Dolez}, N. and {Chevreton}, M. and {Barstow}, M.~A. and {Kepler}, S.~O. and {Giovannini}, O. and {Augusteijn}, T. and {Hansen}, C.~J. and {Kawaler}, S.~D.},
        title = "{Understanding the Cool DA White Dwarf Pulsator, G29-38}",
      journal = {\apj},
         year = 1998,
        month = mar,
       volume = {495},
       number = {1},
        pages = {424-434},
          doi = {10.1086/305259},
archivePrefix = {arXiv},
       eprint = {astro-ph/9711123},
 primaryClass = {astro-ph},
       adsurl = {https://ui.adsabs.harvard.edu/abs/1998ApJ...495..424K}
}

@ARTICLE{2023MNRAS.526.2846U,
       author = {{Uzundag}, Murat and {De Ger{\'o}nimo}, Francisco C. and {C{\'o}rsico}, Alejandro H. and {Silvotti}, Roberto and {Bradley}, Paul A. and {Montgomery}, Michael H. and {Catelan}, M{\'a}rcio and {Toloza}, Odette and {Bell}, Keaton J. and {Kepler}, S.~O. and {Althaus}, Leandro G. and {Kleinman}, Scot J. and {Kilic}, Mukremin and {Mullally}, Susan E. and {G{\"a}nsicke}, Boris T. and {B{\k{a}}kowska}, Karolina and {Barber}, Sam and {Nitta}, Atsuko},
        title = "{Asteroseismological analysis of the polluted ZZ Ceti star G 29 - 38 with TESS}",
      journal = {\mnras},
         year = 2023,
        month = dec,
       volume = {526},
       number = {2},
        pages = {2846-2862},
          doi = {10.1093/mnras/stad2776},
archivePrefix = {arXiv},
       eprint = {2309.04809},
 primaryClass = {astro-ph.SR},
       adsurl = {https://ui.adsabs.harvard.edu/abs/2023MNRAS.526.2846U}
}

@ARTICLE{1991MNRAS.251..673B,
       author = {{Brickhill}, A.~J.},
        title = "{The pulsations of ZZ Ceti stars. III. The driving mechanism.}",
      journal = {\mnras},
         year = 1991,
        month = aug,
       volume = {251},
        pages = {673-680},
          doi = {10.1093/mnras/251.4.673},
       adsurl = {https://ui.adsabs.harvard.edu/abs/1991MNRAS.251..673B}
}

@ARTICLE{1999ApJ...511..904G,
       author = {{Goldreich}, Peter and {Wu}, Yanqin},
        title = "{Gravity Modes in ZZ Ceti Stars. I. Quasi-adiabatic Analysis of Overstability}",
      journal = {\apj},
         year = 1999,
        month = feb,
       volume = {511},
       number = {2},
        pages = {904-915},
          doi = {10.1086/306705},
archivePrefix = {arXiv},
       eprint = {astro-ph/9804305},
 primaryClass = {astro-ph},
       adsurl = {https://ui.adsabs.harvard.edu/abs/1999ApJ...511..904G}
}

@ARTICLE{1999ApJ...519..783W,
       author = {{Wu}, Yanqin and {Goldreich}, Peter},
        title = "{Gravity Modes in ZZ Ceti Stars. II. Eigenvalues and Eigenfunctions}",
      journal = {\apj},
         year = 1999,
        month = jul,
       volume = {519},
       number = {2},
        pages = {783-792},
          doi = {10.1086/307412},
       adsurl = {https://ui.adsabs.harvard.edu/abs/1999ApJ...519..783W}
}

@ARTICLE{2016A&A...585A..22Z,
   author = {{Zong}, W. and {Charpinet}, S. and {Vauclair}, G. and {Giammichele}, N. and 
	{Van Grootel}, V.},
    title = "{Amplitude and frequency variations of oscillation modes in the pulsating DB white dwarf star KIC 08626021. The likely signature of nonlinear resonant mode coupling}",
  journal = {\aap},
archivePrefix = "arXiv",
   eprint = {1510.06884},
 primaryClass = "astro-ph.SR",
     year = 2016,
    month = jan,
   volume = 585,
      eid = {A22},
    pages = {A22},
      doi = {10.1051/0004-6361/201526300},
   adsurl = {http://adsabs.harvard.edu/abs/2016A%26A...585A..22Z}
}

@INPROCEEDINGS{2022afas.confE...1K,
       author = {{Kurtz}, Donald},
        title = "{Asteroseismology across the HR diagram}",
    booktitle = {Annual Conference and General Assembly of the},
         year = 2022,
        month = mar,
          eid = {1},
        pages = {1},
          doi = {10.48550/arXiv.2201.11629},
archivePrefix = {arXiv},
       eprint = {2201.11629},
 primaryClass = {astro-ph.SR},
       adsurl = {https://ui.adsabs.harvard.edu/abs/2022afas.confE...1K}
}

@ARTICLE{2021RvMP...93a5001A,
       author = {{Aerts}, C.},
        title = "{Probing the interior physics of stars through asteroseismology}",
      journal = {Reviews of Modern Physics},
         year = 2021,
        month = jan,
       volume = {93},
       number = {1},
          eid = {015001},
        pages = {015001},
          doi = {10.1103/RevModPhys.93.015001},
archivePrefix = {arXiv},
       eprint = {1912.12300},
 primaryClass = {astro-ph.SR},
       adsurl = {https://ui.adsabs.harvard.edu/abs/2021RvMP...93a5001A}
}

@BOOK{2010aste.bookA,
       author = {{Aerts}, Conny and {Christensen-Dalsgaard}, J{\o}rgen and {Kurtz}, Donald W.},
        title = "{Asteroseismology}",
         year = 2010,
    publisher = {Springer},
          doi = {10.1007/978-1-4020-5803-5},
       adsurl = {https://ui.adsabs.harvard.edu/abs/2010aste.book.....A}
}

@ARTICLE{2013ARAA..51..353C,
       author = {{Chaplin}, William J. and {Miglio}, Andrea},
        title = "{Asteroseismology of Solar-Type and Red-Giant Stars}",
      journal = {\araa},
         year = 2013,
        month = aug,
       volume = {51},
       number = {1},
        pages = {353-392},
          doi = {10.1146/annurev-astro-082812-140938},
archivePrefix = {arXiv},
       eprint = {1303.1957},
 primaryClass = {astro-ph.SR},
       adsurl = {https://ui.adsabs.harvard.edu/abs/2013ARA&A..51..353C}
}

@INPROCEEDINGS{2017ASPC..509..303B,
       author = {{Bell}, K.~J. and {Hermes}, J.~J. and {Montgomery}, M.~H. and {Winget}, D.~E. and {Gentile Fusillo}, N.~P. and {Raddi}, R. and {G{\"a}nsicke}, B.~T.},
        title = "{The First Six Outbursting Cool DA White Dwarf Pulsators}",
    booktitle = {20th European White Dwarf Workshop},
         year = 2017,
       editor = {{Tremblay}, P. -E. and {Gaensicke}, B. and {Marsh}, T.},
       series = {Astronomical Society of the Pacific Conference Series},
       volume = {509},
        month = mar,
        pages = {303},
          doi = {10.48550/arXiv.1609.09097},
archivePrefix = {arXiv},
       eprint = {1609.09097},
 primaryClass = {astro-ph.SR},
       adsurl = {https://ui.adsabs.harvard.edu/abs/2017ASPC..509..303B}
}

@ARTICLE{2012ApJ...748L..10M,
       author = {{Metcalfe}, T.~S. and {Chaplin}, W.~J. and {Appourchaux}, T. and {Garc{\'\i}a}, R.~A. and {Basu}, S. and {Brand{\~a}o}, I. and {Creevey}, O.~L. and {Deheuvels}, S. and {Do{\v{g}}an}, G. and {Eggenberger}, P. and {Karoff}, C. and {Miglio}, A. and {Stello}, D. and {Y{\i}ld{\i}z}, M. and {{\c{C}}elik}, Z. and {Antia}, H.~M. and {Benomar}, O. and {Howe}, R. and {R{\'e}gulo}, C. and {Salabert}, D. and {Stahn}, T. and {Bedding}, T.~R. and {Davies}, G.~R. and {Elsworth}, Y. and {Gizon}, L. and {Hekker}, S. and {Mathur}, S. and {Mosser}, B. and {Bryson}, S.~T. and {Still}, M.~D. and {Christensen-Dalsgaard}, J. and {Gilliland}, R.~L. and {Kawaler}, S.~D. and {Kjeldsen}, H. and {Ibrahim}, K.~A. and {Klaus}, T.~C. and {Li}, J.},
        title = "{Asteroseismology of the Solar Analogs 16 Cyg A and B from Kepler Observations}",
      journal = {\apjl},
         year = 2012,
        month = mar,
       volume = {748},
       number = {1},
          eid = {L10},
        pages = {L10},
          doi = {10.1088/2041-8205/748/1/L10},
archivePrefix = {arXiv},
       eprint = {1201.5966},
 primaryClass = {astro-ph.SR},
       adsurl = {https://ui.adsabs.harvard.edu/abs/2012ApJ...748L..10M}
}

@ARTICLE{2017ApJ...835..172L,
       author = {{Lund}, Mikkel N. and {Silva Aguirre}, V{\'\i}ctor and {Davies}, Guy R. and {Chaplin}, William J. and {Christensen-Dalsgaard}, J{\o}rgen and {Houdek}, G{\"u}nter and {White}, Timothy R. and {Bedding}, Timothy R. and {Ball}, Warrick H. and {Huber}, Daniel and {Antia}, H.~M. and {Lebreton}, Yveline and {Latham}, David W. and {Handberg}, Rasmus and {Verma}, Kuldeep and {Basu}, Sarbani and {Casagrande}, Luca and {Justesen}, Anders B. and {Kjeldsen}, Hans and {Mosumgaard}, Jakob R.},
        title = "{Standing on the Shoulders of Dwarfs: the Kepler Asteroseismic LEGACY Sample. I. Oscillation Mode Parameters}",
      journal = {\apj},
         year = 2017,
        month = feb,
       volume = {835},
       number = {2},
          eid = {172},
        pages = {172},
          doi = {10.3847/1538-4357/835/2/172},
archivePrefix = {arXiv},
       eprint = {1612.00436},
 primaryClass = {astro-ph.SR},
       adsurl = {https://ui.adsabs.harvard.edu/abs/2017ApJ...835..172L}
}

@ARTICLE{2017ApJ...835..173S,
       author = {{Silva Aguirre}, V{\'\i}ctor and {Lund}, Mikkel N. and {Antia}, H.~M. and {Ball}, Warrick H. and {Basu}, Sarbani and {Christensen-Dalsgaard}, J{\o}rgen and {Lebreton}, Yveline and {Reese}, Daniel R. and {Verma}, Kuldeep and {Casagrande}, Luca and {Justesen}, Anders B. and {Mosumgaard}, Jakob R. and {Chaplin}, William J. and {Bedding}, Timothy R. and {Davies}, Guy R. and {Handberg}, Rasmus and {Houdek}, G{\"u}nter and {Huber}, Daniel and {Kjeldsen}, Hans and {Latham}, David W. and {White}, Timothy R. and {Coelho}, Hugo R. and {Miglio}, Andrea and {Rendle}, Ben},
        title = "{Standing on the Shoulders of Dwarfs: the Kepler Asteroseismic LEGACY Sample. II.Radii, Masses, and Ages}",
      journal = {\apj},
         year = 2017,
        month = feb,
       volume = {835},
       number = {2},
          eid = {173},
        pages = {173},
          doi = {10.3847/1538-4357/835/2/173},
archivePrefix = {arXiv},
       eprint = {1611.08776},
 primaryClass = {astro-ph.SR},
       adsurl = {https://ui.adsabs.harvard.edu/abs/2017ApJ...835..173S}
}

@ARTICLE{1968ApJ...153..151L,
       author = {{Landolt}, Arlo U.},
        title = "{A New Short-Period Blue Variable}",
      journal = {\apj},
         year = 1968,
        month = jul,
       volume = {153},
        pages = {151},
          doi = {10.1086/149645},
       adsurl = {https://ui.adsabs.harvard.edu/abs/1968ApJ...153..151L}
}

@ARTICLE{1994ApJ...430..839W,
       author = {{Winget}, D.~E. and {Nather}, R.~E. and {Clemens}, J.~C. and {Provencal}, J.~L. and {Kleinman}, S.~J. and {Bradley}, P.~A. and {Claver}, C.~F. and {Dixson}, J.~S. and {Montgomery}, M.~H. and {Hansen}, C.~J. and {Hine}, B.~P. and {Birch}, P. and {Candy}, M. and {Marar}, T.~M.~K. and {Seetha}, S. and {Ashoka}, B.~N. and {Leibowitz}, E.~M. and {O'Donoghue}, D. and {Warner}, B. and {Buckley}, D.~A.~H. and {Tripe}, P. and {Vauclair}, G. and {Dolez}, N. and {Chevreton}, M. and {Serre}, T. and {Garrido}, R. and {Kepler}, S.~O. and {Kanaan}, A. and {Augusteijn}, T. and {Wood}, M.~A. and {Bergeron}, P. and {Grauer}, A.~D.},
        title = "{Whole Earth Telescope Observations of the DBV White Dwarf GD 358}",
      journal = {\apj},
         year = 1994,
        month = aug,
       volume = {430},
        pages = {839},
          doi = {10.1086/174455},
       adsurl = {https://ui.adsabs.harvard.edu/abs/1994ApJ...430..839W}
}

@ARTICLE{1996A&A...314..182P,
       author = {{Pfeiffer}, B. and {Vauclair}, G. and {Dolez}, N. and {Chevreton}, M. and {Fremy}, J. -R. and {Barstow}, M. and {Belmonte}, J.~A. and {Kepler}, S.~O. and {Kanaan}, A. and {Giovannini}, O. and {Fontaine}, G. and {Bergeron}, P. and {Wesemael}, F. and {Grauer}, A.~D. and {Nather}, R.~E. and {Winget}, D.~E. and {Provencal}, J. and {Clemens}, J.~C. and {Bradley}, P.~A. and {Dixson}, J. and {Kleinman}, S.~J. and {Watson}, T.~K. and {Claver}, C.~F. and {Matzeh}, T. and {Leibowitz}, E.~M. and {Moskalik}, P.},
        title = "{Whole Earth Telescope observations and seismological analysis of the cool ZZ Ceti star GD 154.}",
      journal = {\aap},
         year = 1996,
        month = oct,
       volume = {314},
        pages = {182-190},
       adsurl = {https://ui.adsabs.harvard.edu/abs/1996A&A...314..182P}
}

@ARTICLE{1995ApJ...447..874K,
       author = {{Kepler}, S.~O. and {Giovannini}, O. and {Wood}, M.~A. and {Nather}, R.~E. and {Winget}, D.~E. and {Kanaan}, A. and {Kleinman}, S.~J. and {Bradley}, P.~A. and {Provencal}, J.~L. and {Clemens}, J.~C. and {Claver}, C.~F. and {Watson}, T.~K. and {Yanagida}, K. and {Krisciunas}, K. and {Marar}, T.~M.~K. and {Seetha}, S. and {Ashoka}, B.~N. and {Leibowitz}, E. and {Mendelson}, H. and {Mazeh}, T. and {Moskalik}, P. and {Krzesinski}, J. and {Pajdosz}, G. and {Zola}, S. and {Solheim}, J. -E. and {Emanuelsen}, P. -I. and {Dolez}, N. and {Vauclair}, G. and {Chevreton}, M. and {Fremy}, J. -R. and {Barstow}, M.~A. and {Sansom}, A.~E. and {Tweedy}, R.~W. and {Wickramasinghe}, D.~T. and {Ferrario}, L. and {Sullivan}, D.~J. and {van der Peet}, A.~J. and {Buckley}, D.~A.~H. and {Chen}, A. -L.},
        title = "{Whole Earth Telescope Observations of the DAV White Dwarf G226-29}",
      journal = {\apj},
         year = 1995,
        month = jul,
       volume = {447},
        pages = {874},
          doi = {10.1086/175924},
       adsurl = {https://ui.adsabs.harvard.edu/abs/1995ApJ...447..874K}
}

@ARTICLE{2021ApJ...906....7K,
       author = {{Kepler}, S.~O. and {Winget}, D.~E. and {Vanderbosch}, Zachary P. and {Castanheira}, Barbara Garcia and {Hermes}, J.~J. and {Bell}, Keaton J. and {Mullally}, Fergal and {Romero}, Alejandra D. and {Montgomery}, M.~H. and {DeGennaro}, Steven and {Winget}, Karen I. and {Chandler}, Dean and {Jeffery}, Elizabeth J. and {Fritzen}, Jamile K. and {Williams}, Kurtis A. and {Chote}, Paul and {Zola}, Staszek},
        title = "{The Pulsating White Dwarf G117-B15A: Still the Most Stable Optical Clock Known}",
      journal = {\apj},
         year = 2021,
        month = jan,
       volume = {906},
       number = {1},
          eid = {7},
        pages = {7},
          doi = {10.3847/1538-4357/abc626},
archivePrefix = {arXiv},
       eprint = {2010.16062},
 primaryClass = {astro-ph.SR},
       adsurl = {https://ui.adsabs.harvard.edu/abs/2021ApJ...906....7K}
}

@ARTICLE{2017ApJS..232...23H,
       author = {{Hermes}, J.~J. and {G{\"a}nsicke}, B.~T. and {Kawaler}, Steven D. and {Greiss}, S. and {Tremblay}, P. -E. and {Gentile Fusillo}, N.~P. and {Raddi}, R. and {Fanale}, S.~M. and {Bell}, Keaton J. and {Dennihy}, E. and {Fuchs}, J.~T. and {Dunlap}, B.~H. and {Clemens}, J.~C. and {Montgomery}, M.~H. and {Winget}, D.~E. and {Chote}, P. and {Marsh}, T.~R. and {Redfield}, S.},
        title = "{White Dwarf Rotation as a Function of Mass and a Dichotomy of Mode Line Widths: Kepler Observations of 27 Pulsating DA White Dwarfs through K2 Campaign 8}",
      journal = {\apjs},
         year = 2017,
        month = oct,
       volume = {232},
       number = {2},
          eid = {23},
        pages = {23},
          doi = {10.3847/1538-4365/aa8bb5},
archivePrefix = {arXiv},
       eprint = {1709.07004},
 primaryClass = {astro-ph.SR},
       adsurl = {https://ui.adsabs.harvard.edu/abs/2017ApJS..232...23H}
}

@ARTICLE{2022A&A...663A.167A,
       author = {{Althaus}, Leandro G. and {C{\'o}rsico}, Alejandro H.},
        title = "{New DA white dwarf models for asteroseismology of ZZ Ceti stars}",
      journal = {\aap},
         year = 2022,
        month = jul,
       volume = {663},
          eid = {A167},
        pages = {A167},
          doi = {10.1051/0004-6361/202243943},
archivePrefix = {arXiv},
       eprint = {2205.14126},
 primaryClass = {astro-ph.SR},
       adsurl = {https://ui.adsabs.harvard.edu/abs/2022A&A...663A.167A}
}

@ARTICLE{2008PASP..120.1043F,
       author = {{Fontaine}, G. and {Brassard}, P.},
        title = "{The Pulsating White Dwarf Stars}",
      journal = {\pasp},
         year = 2008,
        month = oct,
       volume = {120},
       number = {872},
        pages = {1043},
          doi = {10.1086/592788},
       adsurl = {https://ui.adsabs.harvard.edu/abs/2008PASP..120.1043F}
}

@misc{2022ascl.soft07007B,
       author = {{Bell}, Keaton},
        title = "{Pyriod: Period detection and fitting routines}",
 howpublished = {Astrophysics Source Code Library, record ascl:2207.007},
         year = 2022,
        month = jul,
          eid = {ascl:2207.007},
archivePrefix = {ascl},
       eprint = {2207.007},
       adsurl = {https://ui.adsabs.harvard.edu/abs/2022ascl.soft07007B}
}

@ARTICLE{2012MNRAS.420.1462R,
       author = {{Romero}, A.~D. and {C{\'o}rsico}, A.~H. and {Althaus}, L.~G. and {Kepler}, S.~O. and {Castanheira}, B.~G. and {Miller Bertolami}, M.~M.},
        title = "{Toward ensemble asteroseismology of ZZ Ceti stars with fully evolutionary models}",
      journal = {\mnras},
         year = 2012,
        month = feb,
       volume = {420},
       number = {2},
        pages = {1462-1480},
          doi = {10.1111/j.1365-2966.2011.20134.x},
archivePrefix = {arXiv},
       eprint = {1109.6682},
 primaryClass = {astro-ph.SR},
       adsurl = {https://ui.adsabs.harvard.edu/abs/2012MNRAS.420.1462R}
}

@ARTICLE{2016ApJS..223...10G,
       author = {{Giammichele}, N. and {Fontaine}, G. and {Brassard}, P. and {Charpinet}, S.},
        title = "{A New Analysis of the Two Classical ZZ Ceti White Dwarfs GD 165 and Ross 548. II. Seismic Modeling}",
      journal = {\apjs},
         year = 2016,
        month = mar,
       volume = {223},
       number = {1},
          eid = {10},
        pages = {10},
          doi = {10.3847/0067-0049/223/1/10},
       adsurl = {https://ui.adsabs.harvard.edu/abs/2016ApJS..223...10G}
}

@ARTICLE{2014MNRAS.438.3086G,
       author = {{Greiss}, S. and {G{\"a}nsicke}, B.~T. and {Hermes}, J.~J. and {Steeghs}, D. and {Koester}, D. and {Ramsay}, G. and {Barclay}, T. and {Townsley}, D.~M.},
        title = "{KIC 11911480: the second ZZ Ceti in the Kepler field}",
      journal = {\mnras},
         year = 2014,
        month = mar,
       volume = {438},
       number = {4},
        pages = {3086-3092},
          doi = {10.1093/mnras/stt2420},
archivePrefix = {arXiv},
       eprint = {1312.4541},
 primaryClass = {astro-ph.SR},
       adsurl = {https://ui.adsabs.harvard.edu/abs/2014MNRAS.438.3086G}
}

@ARTICLE{2020ApJ...890...11M,
       author = {{Montgomery}, M.~H. and {Hermes}, J.~J. and {Winget}, D.~E. and {Dunlap}, B.~H. and {Bell}, K.~J.},
        title = "{Limits on Mode Coherence in Pulsating DA White Dwarfs Due to a Nonstatic Convection Zone}",
      journal = {\apj},
         year = 2020,
        month = feb,
       volume = {890},
       number = {1},
          eid = {11},
        pages = {11},
          doi = {10.3847/1538-4357/ab6a0e},
archivePrefix = {arXiv},
       eprint = {2001.05048},
 primaryClass = {astro-ph.SR},
       adsurl = {https://ui.adsabs.harvard.edu/abs/2020ApJ...890...11M}
}

@ARTICLE{2013A&A...558A..33A,
       author = {{Astropy Collaboration} and {Robitaille}, Thomas P. and {Tollerud}, Erik J. and {Greenfield}, Perry and {Droettboom}, Michael and {Bray}, Erik and {Aldcroft}, Tom and {Davis}, Matt and {Ginsburg}, Adam and {Price-Whelan}, Adrian M. and {Kerzendorf}, Wolfgang E. and {Conley}, Alexander and {Crighton}, Neil and {Barbary}, Kyle and {Muna}, Demitri and {Ferguson}, Henry and {Grollier}, Fr{\'e}d{\'e}ric and {Parikh}, Madhura M. and {Nair}, Prasanth H. and {Unther}, Hans M. and {Deil}, Christoph and {Woillez}, Julien and {Conseil}, Simon and {Kramer}, Roban and {Turner}, James E.~H. and {Singer}, Leo and {Fox}, Ryan and {Weaver}, Benjamin A. and {Zabalza}, Victor and {Edwards}, Zachary I. and {Azalee Bostroem}, K. and {Burke}, D.~J. and {Casey}, Andrew R. and {Crawford}, Steven M. and {Dencheva}, Nadia and {Ely}, Justin and {Jenness}, Tim and {Labrie}, Kathleen and {Lim}, Pey Lian and {Pierfederici}, Francesco and {Pontzen}, Andrew and {Ptak}, Andy and {Refsdal}, Brian and {Servillat}, Mathieu and {Streicher}, Ole},
        title = "{Astropy: A community Python package for astronomy}",
      journal = {\aap},
         year = 2013,
        month = oct,
       volume = {558},
          eid = {A33},
        pages = {A33},
          doi = {10.1051/0004-6361/201322068},
archivePrefix = {arXiv},
       eprint = {1307.6212},
 primaryClass = {astro-ph.IM},
       adsurl = {https://ui.adsabs.harvard.edu/abs/2013A&A...558A..33A}
}

@ARTICLE{2018AJ....156..123A,
       author = {{Astropy Collaboration} and {Price-Whelan}, A.~M. and
         {Sip{\H{o}}cz}, B.~M. and {G{\"u}nther}, H.~M. and {Lim}, P.~L. and
         {Crawford}, S.~M. and {Conseil}, S. and {Shupe}, D.~L. and
         {Craig}, M.~W. and {Dencheva}, N. and {Ginsburg}, A. and {Vand
        erPlas}, J.~T. and {Bradley}, L.~D. and {P{\'e}rez-Su{\'a}rez}, D. and
         {de Val-Borro}, M. and {Aldcroft}, T.~L. and {Cruz}, K.~L. and
         {Robitaille}, T.~P. and {Tollerud}, E.~J. and {Ardelean}, C. and
         {Babej}, T. and {Bach}, Y.~P. and {Bachetti}, M. and {Bakanov}, A.~V. and
         {Bamford}, S.~P. and {Barentsen}, G. and {Barmby}, P. and
         {Baumbach}, A. and {Berry}, K.~L. and {Biscani}, F. and {Boquien}, M. and
         {Bostroem}, K.~A. and {Bouma}, L.~G. and {Brammer}, G.~B. and
         {Bray}, E.~M. and {Breytenbach}, H. and {Buddelmeijer}, H. and
         {Burke}, D.~J. and {Calderone}, G. and {Cano Rodr{\'\i}guez}, J.~L. and
         {Cara}, M. and {Cardoso}, J.~V.~M. and {Cheedella}, S. and {Copin}, Y. and
         {Corrales}, L. and {Crichton}, D. and {D'Avella}, D. and {Deil}, C. and
         {Depagne}, {\'E}. and {Dietrich}, J.~P. and {Donath}, A. and
         {Droettboom}, M. and {Earl}, N. and {Erben}, T. and {Fabbro}, S. and
         {Ferreira}, L.~A. and {Finethy}, T. and {Fox}, R.~T. and
         {Garrison}, L.~H. and {Gibbons}, S.~L.~J. and {Goldstein}, D.~A. and
         {Gommers}, R. and {Greco}, J.~P. and {Greenfield}, P. and
         {Groener}, A.~M. and {Grollier}, F. and {Hagen}, A. and {Hirst}, P. and
         {Homeier}, D. and {Horton}, A.~J. and {Hosseinzadeh}, G. and {Hu}, L. and
         {Hunkeler}, J.~S. and {Ivezi{\'c}}, {\v{Z}}. and {Jain}, A. and
         {Jenness}, T. and {Kanarek}, G. and {Kendrew}, S. and {Kern}, N.~S. and
         {Kerzendorf}, W.~E. and {Khvalko}, A. and {King}, J. and {Kirkby}, D. and
         {Kulkarni}, A.~M. and {Kumar}, A. and {Lee}, A. and {Lenz}, D. and
         {Littlefair}, S.~P. and {Ma}, Z. and {Macleod}, D.~M. and
         {Mastropietro}, M. and {McCully}, C. and {Montagnac}, S. and
         {Morris}, B.~M. and {Mueller}, M. and {Mumford}, S.~J. and {Muna}, D. and
         {Murphy}, N.~A. and {Nelson}, S. and {Nguyen}, G.~H. and
         {Ninan}, J.~P. and {N{\"o}the}, M. and {Ogaz}, S. and {Oh}, S. and
         {Parejko}, J.~K. and {Parley}, N. and {Pascual}, S. and {Patil}, R. and
         {Patil}, A.~A. and {Plunkett}, A.~L. and {Prochaska}, J.~X. and
         {Rastogi}, T. and {Reddy Janga}, V. and {Sabater}, J. and
         {Sakurikar}, P. and {Seifert}, M. and {Sherbert}, L.~E. and
         {Sherwood-Taylor}, H. and {Shih}, A.~Y. and {Sick}, J. and
         {Silbiger}, M.~T. and {Singanamalla}, S. and {Singer}, L.~P. and
         {Sladen}, P.~H. and {Sooley}, K.~A. and {Sornarajah}, S. and
         {Streicher}, O. and {Teuben}, P. and {Thomas}, S.~W. and
         {Tremblay}, G.~R. and {Turner}, J.~E.~H. and {Terr{\'o}n}, V. and
         {van Kerkwijk}, M.~H. and {de la Vega}, A. and {Watkins}, L.~L. and
         {Weaver}, B.~A. and {Whitmore}, J.~B. and {Woillez}, J. and
         {Zabalza}, V. and {Astropy Contributors}},
        title = "{The Astropy Project: Building an Open-science Project and Status of the v2.0 Core Package}",
      journal = {\aj},
         year = "2018",
        month = "Sep",
       volume = {156},
       number = {3},
          eid = {123},
        pages = {123},
          doi = {10.3847/1538-3881/aabc4f},
archivePrefix = {arXiv},
       eprint = {1801.02634},
 primaryClass = {astro-ph.IM},
       adsurl = {https://ui.adsabs.harvard.edu/abs/2018AJ....156..123A}
}

@Article{Hunter:2007,
  Author    = {Hunter, J. D.},
  Title     = {Matplotlib: A 2D graphics environment},
  Journal   = {Computing in Science \& Engineering},
  Volume    = {9},
  Number    = {3},
  Pages     = {90--95},
  publisher = {IEEE COMPUTER SOC},
  doi       = {10.1109/MCSE.2007.55},
  year      = 2007
}

@Article{         harris2020array,
 title         = {Array programming with {NumPy}},
 author        = {Charles R. Harris and K. Jarrod Millman and St{\'{e}}fan J.
                 van der Walt and Ralf Gommers and Pauli Virtanen and David
                 Cournapeau and Eric Wieser and Julian Taylor and Sebastian
                 Berg and Nathaniel J. Smith and Robert Kern and Matti Picus
                 and Stephan Hoyer and Marten H. van Kerkwijk and Matthew
                 Brett and Allan Haldane and Jaime Fern{\'{a}}ndez del
                 R{\'{i}}o and Mark Wiebe and Pearu Peterson and Pierre
                 G{\'{e}}rard-Marchant and Kevin Sheppard and Tyler Reddy and
                 Warren Weckesser and Hameer Abbasi and Christoph Gohlke and
                 Travis E. Oliphant},
 year          = {2020},
 month         = sep,
 journal       = {Nature},
 volume        = {585},
 number        = {7825},
 pages         = {357--362},
 doi           = {10.1038/s41586-020-2649-2},
 publisher     = {Springer Science and Business Media {LLC}},
 url           = {https://doi.org/10.1038/s41586-020-2649-2}
}

@misc{pandas2020,
    author       = {The pandas development team},
    title        = {pandas-dev/pandas: Pandas},
    month        = feb,
    year         = 2020,
    publisher    = {Zenodo},
    version      = {latest},
    doi          = {10.5281/zenodo.3509134},
    url          = {https://doi.org/10.5281/zenodo.3509134}
}

@ARTICLE{2020SciPy-NMeth,
  author  = {Virtanen, Pauli and Gommers, Ralf and Oliphant, Travis E. and
            Haberland, Matt and Reddy, Tyler and Cournapeau, David and
            Burovski, Evgeni and Peterson, Pearu and Weckesser, Warren and
            Bright, Jonathan and {van der Walt}, St{\'e}fan J. and
            Brett, Matthew and Wilson, Joshua and Millman, K. Jarrod and
            Mayorov, Nikolay and Nelson, Andrew R. J. and Jones, Eric and
            Kern, Robert and Larson, Eric and Carey, C J and
            Polat, {\.I}lhan and Feng, Yu and Moore, Eric W. and
            {VanderPlas}, Jake and Laxalde, Denis and Perktold, Josef and
            Cimrman, Robert and Henriksen, Ian and Quintero, E. A. and
            Harris, Charles R. and Archibald, Anne M. and
            Ribeiro, Ant{\^o}nio H. and Pedregosa, Fabian and
            {van Mulbregt}, Paul and {SciPy 1.0 Contributors}},
  title   = {{{SciPy} 1.0: Fundamental Algorithms for Scientific
            Computing in Python}},
  journal = {Nature Methods},
  year    = {2020},
  volume  = {17},
  pages   = {261--272},
  adsurl  = {https://rdcu.be/b08Wh},
  doi     = {10.1038/s41592-019-0686-2},
}

@ARTICLE{2005CoAst.146...53L,
       author = {{Lenz}, P. and {Breger}, M.},
        title = "{Period04 User Guide}",
      journal = {Communications in Asteroseismology},
         year = 2005,
        month = jun,
       volume = {146},
        pages = {53-136},
          doi = {10.1553/cia146s53},
       adsurl = {https://ui.adsabs.harvard.edu/abs/2005CoAst.146...53L}
}

@ARTICLE{2014ApJ...789...85H,
       author = {{Hermes}, J.~J. and {Charpinet}, S. and {Barclay}, Thomas and {Pak{\v{s}}tien{\.{e}}}, E. and {Mullally}, Fergal and {Kawaler}, Steven D. and {Bloemen}, S. and {Castanheira}, Barbara G. and {Winget}, D.~E. and {Montgomery}, M.~H. and {Van Grootel}, V. and {Huber}, Daniel and {Still}, Martin and {Howell}, Steve B. and {Caldwell}, Douglas A. and {Haas}, Michael R. and {Bryson}, Stephen T.},
        title = "{Precision Asteroseismology of the Pulsating White Dwarf GD 1212 Using a Two-wheel-controlled Kepler Spacecraft}",
      journal = {\apj},
         year = 2014,
        month = jul,
       volume = {789},
       number = {1},
          eid = {85},
        pages = {85},
          doi = {10.1088/0004-637X/789/1/85},
archivePrefix = {arXiv},
       eprint = {1405.3665},
 primaryClass = {astro-ph.SR},
       adsurl = {https://ui.adsabs.harvard.edu/abs/2014ApJ...789...85H}
}

@misc{2012ascl.soft08004S,
       author = {{Still}, Martin and {Barclay}, Tom},
        title = "{PyKE: Reduction and analysis of Kepler Simple Aperture Photometry data}",
 howpublished = {Astrophysics Source Code Library, record ascl:1208.004},
         year = 2012,
        month = aug,
          eid = {ascl:1208.004},
       adsurl = {https://ui.adsabs.harvard.edu/abs/2012ascl.soft08004S}
}

@ARTICLE{2014PASP..126..948V,
       author = {{Vanderburg}, Andrew and {Johnson}, John Asher},
        title = "{A Technique for Extracting Highly Precise Photometry for the Two-Wheeled Kepler Mission}",
      journal = {\pasp},
         year = 2014,
        month = oct,
       volume = {126},
       number = {944},
        pages = {948},
          doi = {10.1086/678764},
archivePrefix = {arXiv},
       eprint = {1408.3853},
 primaryClass = {astro-ph.IM},
       adsurl = {https://ui.adsabs.harvard.edu/abs/2014PASP..126..948V}
}

@ARTICLE{2015ApJ...810L...5H,
       author = {{Hermes}, J.~J. and {Montgomery}, M.~H. and {Bell}, Keaton J. and {Chote}, P. and {G{\"a}nsicke}, B.~T. and {Kawaler}, Steven D. and {Clemens}, J.~C. and {Dunlap}, Bart H. and {Winget}, D.~E. and {Armstrong}, D.~J.},
        title = "{A Second Case of Outbursts in a Pulsating White Dwarf Observed by Kepler}",
      journal = {\apjl},
         year = 2015,
        month = sep,
       volume = {810},
       number = {1},
          eid = {L5},
        pages = {L5},
          doi = {10.1088/2041-8205/810/1/L5},
archivePrefix = {arXiv},
       eprint = {1507.06319},
 primaryClass = {astro-ph.SR},
       adsurl = {https://ui.adsabs.harvard.edu/abs/2015ApJ...810L...5H}
}

@ARTICLE{2016ApJ...829...82B,
       author = {{Bell}, Keaton J. and {Hermes}, J.~J. and {Montgomery}, M.~H. and {Gentile Fusillo}, N.~P. and {Raddi}, R. and {G{\"a}nsicke}, B.~T. and {Winget}, D.~E. and {Dennihy}, E. and {Gianninas}, A. and {Tremblay}, P. -E. and {Chote}, P. and {Winget}, K.~I.},
        title = "{Outbursts in Two New Cool Pulsating DA White Dwarfs}",
      journal = {\apj},
         year = 2016,
        month = oct,
       volume = {829},
       number = {2},
          eid = {82},
        pages = {82},
          doi = {10.3847/0004-637X/829/2/82},
archivePrefix = {arXiv},
       eprint = {1607.01392},
 primaryClass = {astro-ph.SR},
       adsurl = {https://ui.adsabs.harvard.edu/abs/2016ApJ...829...82B}
}

@ARTICLE{2015ApJ...809...14B,
       author = {{Bell}, Keaton J. and {Hermes}, J.~J. and {Bischoff-Kim}, A. and {Moorhead}, Sean and {Montgomery}, M.~H. and {{\O}stensen}, Roy and {Castanheira}, Barbara G. and {Winget}, D.~E.},
        title = "{KIC 4552982: Outbursts and Asteroseismology from the Longest Pseudo-continuous Light Curve of a ZZ Ceti}",
      journal = {\apj},
         year = 2015,
        month = aug,
       volume = {809},
       number = {1},
          eid = {14},
        pages = {14},
          doi = {10.1088/0004-637X/809/1/14},
archivePrefix = {arXiv},
       eprint = {1506.07878},
 primaryClass = {astro-ph.SR},
       adsurl = {https://ui.adsabs.harvard.edu/abs/2015ApJ...809...14B}
}

@ARTICLE{2001ApJ...546..469W,
       author = {{Wu}, Yanqin and {Goldreich}, Peter},
        title = "{Gravity Modes in ZZ Ceti Stars. IV. Amplitude Saturation by Parametric Instability}",
      journal = {\apj},
         year = 2001,
        month = jan,
       volume = {546},
       number = {1},
        pages = {469-483},
          doi = {10.1086/318234},
archivePrefix = {arXiv},
       eprint = {astro-ph/0003163},
 primaryClass = {astro-ph},
       adsurl = {https://ui.adsabs.harvard.edu/abs/2001ApJ...546..469W}
}

@ARTICLE{2024MNRAS.527.8687O,
       author = {{O'Brien}, Mairi W. and {Tremblay}, P. -E. and {Klein}, B.~L. and {Koester}, D. and {Melis}, C. and {B{\'e}dard}, A. and {Cukanovaite}, E. and {Cunningham}, T. and {Doyle}, A.~E. and {G{\"a}nsicke}, B.~T. and {Gentile Fusillo}, N.~P. and {Hollands}, M.~A. and {McCleery}, J. and {Pelisoli}, I. and {Toonen}, S. and {Weinberger}, A.~J. and {Zuckerman}, B.},
        title = "{The 40 pc sample of white dwarfs from Gaia}",
      journal = {\mnras},
         year = 2024,
        month = jan,
       volume = {527},
       number = {3},
        pages = {8687-8705},
          doi = {10.1093/mnras/stad3773},
archivePrefix = {arXiv},
       eprint = {2312.02735},
 primaryClass = {astro-ph.SR},
       adsurl = {https://ui.adsabs.harvard.edu/abs/2024MNRAS.527.8687O}
}

@ARTICLE{2013A&A...559A.104T,
       author = {{Tremblay}, P. -E. and {Ludwig}, H. -G. and {Steffen}, M. and {Freytag}, B.},
        title = "{Spectroscopic analysis of DA white dwarfs with 3D model atmospheres}",
      journal = {\aap},
         year = 2013,
        month = nov,
       volume = {559},
          eid = {A104},
        pages = {A104},
          doi = {10.1051/0004-6361/201322318},
archivePrefix = {arXiv},
       eprint = {1309.0886},
 primaryClass = {astro-ph.SR},
       adsurl = {https://ui.adsabs.harvard.edu/abs/2013A&A...559A.104T}
}

@ARTICLE{2011ApJ...743..138G,
       author = {{Gianninas}, A. and {Bergeron}, P. and {Ruiz}, M.~T.},
        title = "{A Spectroscopic Survey and Analysis of Bright, Hydrogen-rich White Dwarfs}",
      journal = {\apj},
         year = 2011,
        month = dec,
       volume = {743},
       number = {2},
          eid = {138},
        pages = {138},
          doi = {10.1088/0004-637X/743/2/138},
archivePrefix = {arXiv},
       eprint = {1109.3171},
 primaryClass = {astro-ph.SR},
       adsurl = {https://ui.adsabs.harvard.edu/abs/2011ApJ...743..138G}
}

@ARTICLE{2006AJ....132..831G,
       author = {{Gianninas}, A. and {Bergeron}, P. and {Fontaine}, G.},
        title = "{Mapping the ZZ Ceti Instability Strip: Discovery of Six New Pulsators}",
      journal = {\aj},
         year = 2006,
        month = aug,
       volume = {132},
       number = {2},
        pages = {831-835},
          doi = {10.1086/506516},
archivePrefix = {arXiv},
       eprint = {astro-ph/0606135},
 primaryClass = {astro-ph},
       adsurl = {https://ui.adsabs.harvard.edu/abs/2006AJ....132..831G}
}

@ARTICLE{2018ApJ...863...82L,
       author = {{Luan}, Jing and {Goldreich}, Peter},
        title = "{DAVs: Red Edge and Outbursts}",
      journal = {\apj},
         year = 2018,
        month = aug,
       volume = {863},
       number = {1},
          eid = {82},
        pages = {82},
          doi = {10.3847/1538-4357/aad0f4},
archivePrefix = {arXiv},
       eprint = {1711.06367},
 primaryClass = {astro-ph.SR},
       adsurl = {https://ui.adsabs.harvard.edu/abs/2018ApJ...863...82L}
}

@ARTICLE{2006AJ....132.1221H,
       author = {{Holberg}, J.~B. and {Bergeron}, Pierre},
        title = "{Calibration of Synthetic Photometry Using DA White Dwarfs}",
      journal = {\aj},
         year = 2006,
        month = sep,
       volume = {132},
       number = {3},
        pages = {1221-1233},
          doi = {10.1086/505938},
       adsurl = {https://ui.adsabs.harvard.edu/abs/2006AJ....132.1221H}
}

@ARTICLE{WDEC,
       author = {{Bischoff-Kim}, Agn{\`e}s and {Montgomery}, Michael H.},
        title = "{WDEC: A Code for Modeling White Dwarf Structure and Pulsations}",
      journal = {\aj},
         year = 2018,
        month = may,
       volume = {155},
       number = {5},
          eid = {187},
        pages = {187},
          doi = {10.3847/1538-3881/aab70e},
archivePrefix = {arXiv},
       eprint = {1803.03848},
 primaryClass = {astro-ph.IM},
       adsurl = {https://ui.adsabs.harvard.edu/abs/2018AJ....155..187B}
}

@ARTICLE{WET,
       author = {{Nather}, R.~E. and {Winget}, D.~E. and {Clemens}, J.~C. and {Hansen}, C.~J. and {Hine}, B.~P.},
        title = "{The Whole Earth Telescope: A New Astronomical Instrument}",
      journal = {\apj},
         year = 1990,
        month = sep,
       volume = {361},
        pages = {309},
          doi = {10.1086/169196},
       adsurl = {https://ui.adsabs.harvard.edu/abs/1990ApJ...361..309N}
}

@ARTICLE{2011ApJS..192....3P,
       author = {{Paxton}, Bill and {Bildsten}, Lars and {Dotter}, Aaron and {Herwig}, Falk and {Lesaffre}, Pierre and {Timmes}, Frank},
        title = "{Modules for Experiments in Stellar Astrophysics (MESA)}",
      journal = {\apjs},
         year = 2011,
        month = jan,
       volume = {192},
       number = {1},
          eid = {3},
        pages = {3},
          doi = {10.1088/0067-0049/192/1/3},
archivePrefix = {arXiv},
       eprint = {1009.1622},
 primaryClass = {astro-ph.SR},
       adsurl = {https://ui.adsabs.harvard.edu/abs/2011ApJS..192....3P}
}

@ARTICLE{2013MNRAS.435.3406T,
       author = {{Townsend}, R.~H.~D. and {Teitler}, S.~A.},
        title = "{GYRE: an open-source stellar oscillation code based on a new Magnus Multiple Shooting scheme}",
      journal = {\mnras},
         year = 2013,
        month = nov,
       volume = {435},
       number = {4},
        pages = {3406-3418},
          doi = {10.1093/mnras/stt1533},
archivePrefix = {arXiv},
       eprint = {1308.2965},
 primaryClass = {astro-ph.SR},
       adsurl = {https://ui.adsabs.harvard.edu/abs/2013MNRAS.435.3406T}
}
\bibliographystyle{aasjournal}

\appendix

\section{Constraining ``Failed Outburst'' Mode Changes}\label{app:wdec}

We use the White Dwarf Evolution Code (WDEC, \citealt{WDEC}) to explore how the correlated shifts in frequency could be caused by a change in the temperature of the outer layers of the star. WDEC relaxes the structures of parameterized white dwarf models and computes their pulsation frequencies. The challenge with modeling a superficial heating event with a time-independent code like WDEC is that changing the effective temperature (\teff) also changes the temperature at much deeper layers and thus affects all theoretical mode frequencies. We expect a superficial heating event to temporarily thin the outer convection zone until the excess heat is radiated away. We can mimic such an effect with WDEC by modifying the mixing-length parameter ($\alpha$), which characterizes convective efficiency---a lower value of $\alpha$ will produce a thinner convection zone. Our approach to simulating the ``failed'' outburst by modifying the depth of the convection zone via $\alpha$ as a proxy for superficial heating is similar to the approach used to explore interactions with the convection zone as an explanation for the incoherence of long-period pulsations in white dwarfs by \citet[][]{2020ApJ...890...11M}. 

We record the size of the outer convection zone in the WDEC models as $M_{\rm CZ}$, defined as the mass above the location where the Brunt--V\"ais\"al\"a frequency ($N^2$) drops sharply to zero at the base of the convection zone. We calculate sets of models with different values of $\alpha$ and \teff\ to study how the depth of the convection zone responds. These models explore departures from a reference model with mass $M_\star=0.62$\,\msun, \teff\ = $10{,}970$\,K, $\log(M_{\text{env}}/M_{\star}) = -1.703$, 
$\log(M_{\text{He}}/M_{\star}) = -2.186$, 
$\log(M_{\text{H}}/M_{\star}) = -3.998$, and $\alpha = 0.96$, which is qualitatively very similar to the best-fit asteroseismic model of GD 1212 found by \citet{2017ApJ...851...60R}. Figure~\ref{fig:failed_outburst} shows how the mass in the convection zone of the models changes with $\alpha$ and \teff. The apparent stair-step behavior of the individual models (circles) is due to the limited spatial resolution in the outer layers of the models, and we capture the overall trends by fitting polynomials (solid curves).

\begin{figure*}[b!]
    \centering
    {\includegraphics[width=0.55\textwidth]{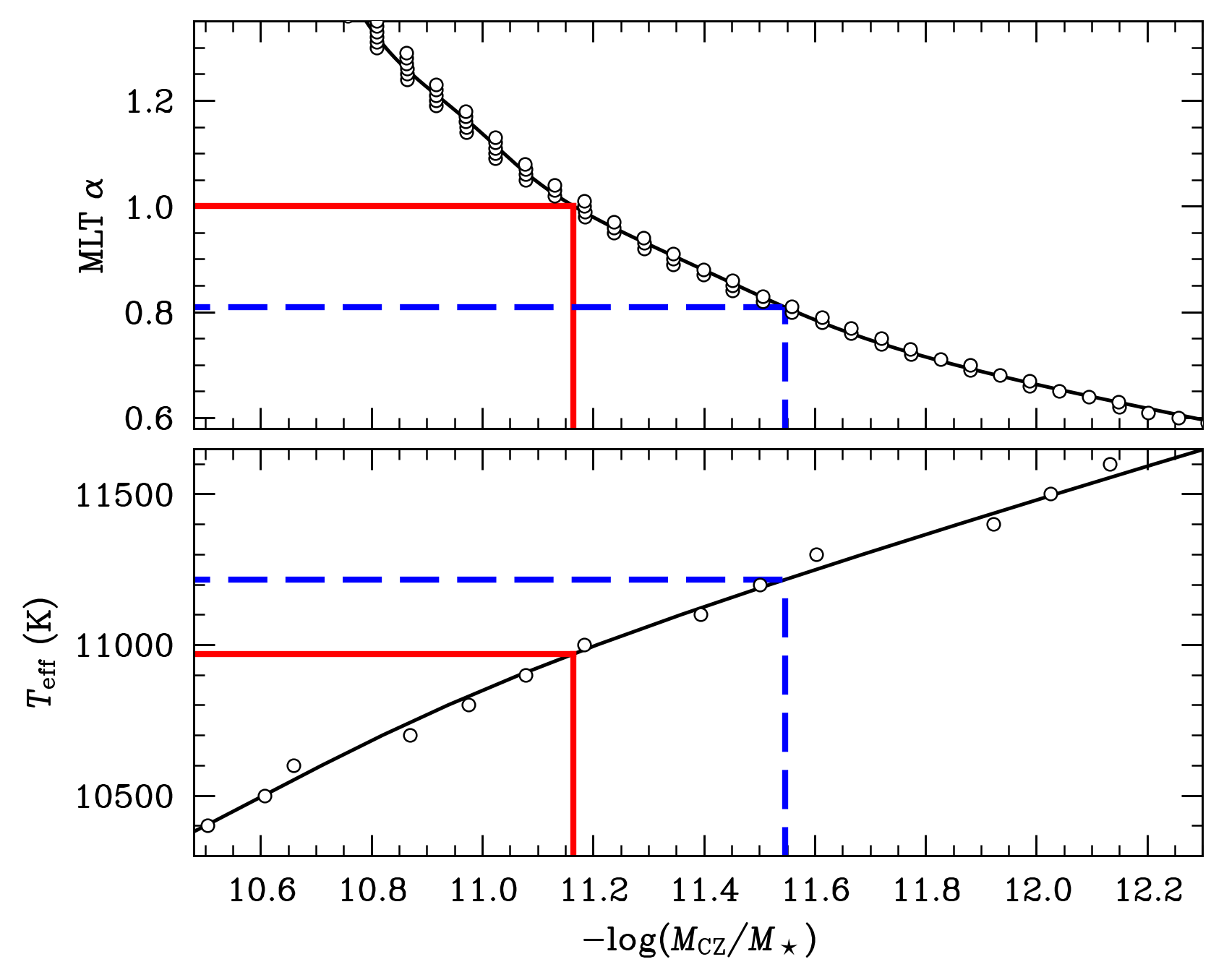}}
    \caption{\textbf{Top Panel:} The location of the base of the convection zone, in terms of the logarithmic mass fraction in the outer convection zone, changes for different values of convective efficiency via the mixing-length parameter $\alpha$. Starting from $\alpha = 1.00$ we find that mode frequencies can change by 5\,\muhz\ if we decrease $\alpha$ by 0.191 to the value marked in blue. This decreases the mass in the convection zone by 0.38 dex. \textbf{Bottom Panel:} We map the effective temperature change needed to move the base of the convection zone to match the amount from the top subplot: in this case $\sim250$\,K.}%
    \label{fig:failed_outburst}
\end{figure*}

To connect the observed frequency shifts in the failed outburst to changes in the convection zone, we compute the numerical rate of change of all $\ell = 1$ and $\ell = 2$ mode frequencies computed with WDEC versus $\alpha$;
we note that this same approach to modeling the change in depth of the convection zone by a change in $\alpha$ was employed by \citet{2020ApJ...890...11M}.
Figure~\ref{fig:failed_outburst} allows us to translate from $\alpha$ (which we use as a proxy for surface heating) to \teff\ at the photosphere: changing $\alpha$ allows us to model a change in the depth of the convection zone (in terms of logarithmic mass fraction), and we interpret this as a change in \teff\ at the photosphere that would produce an equivalent change in the convection-zone depth. We multiply the numerical values of $d\nu/d\alpha$ by $d\alpha/dT_{\text{eff}}$ evaluated near the fiducial model to estimate $d\nu/dT_{\text{eff}}$ for each mode, which are plotted for $\ell=1$ modes in Figure~\ref{fig:df_dTeff_l=1} and for $\ell=2$ modes in Figure~\ref{fig:df_dTeff_l=2}.

The observed modes traced in Figure~\ref{fig:threefailed} show a trend of larger frequency shifts for lower-frequency modes. The slope of this trend matches our modeling results for $\ell=1$ modes (Figure~\ref{fig:df_dTeff_l=1}), while the $\ell=2$ modes in the observed frequency range exhibit the opposite trend (Figure~\ref{fig:df_dTeff_l=2}), with higher-frequency modes changing more rapidly. Because $\ell=2$ modes of comparable radial order occur at higher frequencies, the trend is shifted in frequency relative to the $\ell=1$ case. The scale of $d\nu/dT_{\rm eff}$ is also larger for $\ell=2$ ($\approx 0.03\,\mu{\rm Hz\, K}^{-1}$) than $\ell=1$ ($\approx 0.02\,\mu{\rm Hz\, K}^{-1}$) modes at these frequencies. In addition, we independently calculated frequency changes using results from \citet{2020ApJ...890...11M} that were computed using the MESA stellar evolution code \citep{2011ApJS..192....3P} and the GYRE oscillation code \citep{2013MNRAS.435.3406T}; the frequencies show the same trends as those computed with WDEC and the frequency shifts are similar in magnitude.

Figure~\ref{fig:failed_outburst} demonstrates how the observed frequency shifts of modes are interpreted through our models. Starting at the previously reported spectroscopic \teff\ $= 10{,}970$\,K, the polynomial fitted to the WDEC models estimates a starting convection zone depth of $\log(M_{\rm CZ}/M_\star) = -11.1648$ (solid red lines). From the polynomial for convection zone depth versus $\alpha$, this corresponds to a starting value of $\alpha = 1.000$ (solid red).\footnote{The slight difference from $\alpha = 0.96$ used in the reference models is due to numerical noise in the modeling that we have smoothed over. This does not have an appreciable effect on our inferred change in \teff, which depends on the derivative of the fitted curve in Figure~\ref{fig:failed_outburst} evaluated at reasonable values for this target.} 
The observed amplitude of frequency shifts (5.74\,\muhz\ for the mode at 837\,\muhz, 4.70\,\muhz\ at 937\,\muhz, and 4.22\,\muhz\ at 1089\,\muhz; Figure~\ref{fig:threefailed}) can be reproduced for $\ell=1$ modes by decreasing $\alpha$ by $0.191$, causing $\log(M_{\rm CZ}/M_\star)$ to decrease to $-11.546$ (dashed blue line; a 0.38 dex decrease). This decrease in convection zone depth is equivalent to an increase in \teff\ of $\sim 250$\,K (bottom panel of Figure~\ref{fig:failed_outburst}).

\begin{figure*}[t!]
    \centering
    {\includegraphics[width=0.55\textwidth]{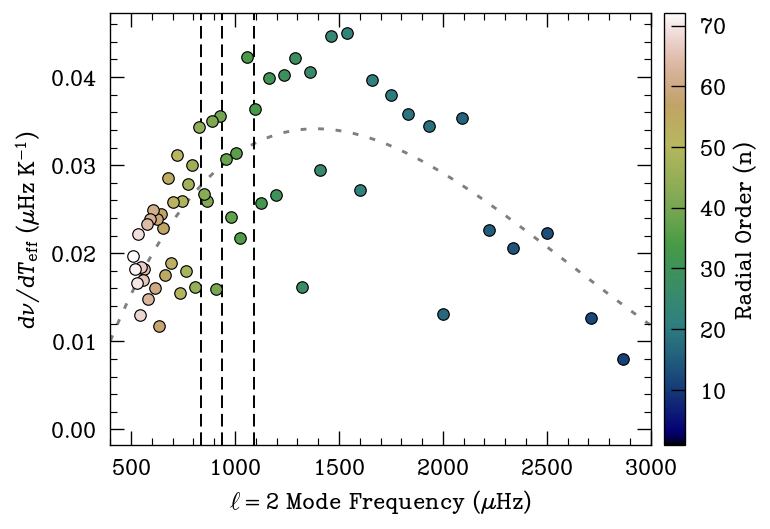}}
    \caption{Following Figure~\ref{fig:df_dTeff_l=1}, we use WDEC to model the expected frequency change as a function of effective temperature for different $\ell=2$ modes in a representative model for \tar. A fifth-order polynomial is shown as a dotted line in the background as a rough analytic estimate. The three independent pulsation modes which undergo a 5\,\muhz\ increase in frequency during the ``failed'' outburst are labeled with dashed black vertical lines. There exists a critical frequency distinguishing modes with greater sensitivity to frequency gradients in effective temperature from those with less sensitivity.}
    \label{fig:df_dTeff_l=2}
\end{figure*}

\section{Observed Frequency Census}
\label{app:periods}

We highlight in Table~\ref{tab:appendix_21d} the full census of modes above the significance threshold in the first 21 days of K2 Campaign 12 observations of \tar\ shown in Figure~\ref{fig:staticFT}. We perform a Lomb-Scargle periodogram analysis to find all signals that exceed $5.4\times$ the mean amplitude of the periodogram (S/N $>$ 5.4) in a window from $500-3000$\,\muhz\  \citep{2015MNRAS.448L..16B}. We perform this analysis using Pyriod \citep{2022ascl.soft07007B}, iteratively prewhitening by the highest-amplitude signals; we prewhiten by fitting a sinusoid with the frequency fixed at this highest amplitude in the periodogram. Once a signal is found, we ignore any other signals within 2\,\muhz, and iteratively prewhiten until all significant signals are found. We use these frequencies and amplitudes as initial guesses for a nonlinear least-squares fit to the 21-day light curve, and report the results in Table~\ref{tab:appendix_21d}. These least-squares results (and reported errors) have significant systematic uncertainties, given the empirical mode instability (e.g., \citealt{2017ApJS..232...23H}). For a handful of modes described in Figure~\ref{fig:multiplets} and Table~\ref{tab:id1}, we relax the S/N threshold because these mode frequencies fit the observed patterns in the rotational splittings. Two modes are marked with a $\dagger$ --- $f_{3b}$ and $f_{3c}$ --- as they are refit from K2 Campaign 0, to help with the identification of the third $\ell=1$ multiplet in Table~\ref{tab:id1}. We exclude $f_{3a}$, $f_{4a}$, and $f_{5a}$ in the calculation of splittings in Table~\ref{tab:id1}.

\begin{table}
\caption{Observed pulsation modes for first 21 days of K2 Campaign 12 observations of \tar.}
\label{tab:appendix_21d}
\centering
\begin{tabular*}{\linewidth}{@{\extracolsep{\fill}}cccccccc}
\hline\hline
ID & Frequency & Period & Amplitude & S/N & Splitting & $\ell$ & $m$ \\
   & ($\mu$Hz) & (s)      & (ppt)     &  & ($\mu$Hz) &  & \\
\hline
$f_{1a}$ & 2687.02  $\pm$ 0.15 &  & 0.049 & 1.7 & 8.67  $\pm$ 0.21 & 1 & $-1$ \\
$f_{1b}$ & 2695.68  $\pm$ 0.14 & 371.0 & 0.051 & 1.8 & 8.67  $\pm$ 0.15 & 1 & 0 \\
$f_{1c}$ & 2704.349  $\pm$ 0.026 &  & 0.281 & 9.8 &    & 1 & $+1$ \\
$f_{2a}$ & 2438.463  $\pm$ 0.096 &  & 0.076 & 2.7 & 1.23  $\pm$ 0.19 & 1 &  \\
$f_{2b}$ & 2439.70  $\pm$ 0.17 &  & 0.043 & 1.5 & 8.59  $\pm$ 0.22 & 1 & $-1$ \\
$f_{2c}$ & 2448.29  $\pm$ 0.13 & 408.4 & 0.055 & 1.9 & 8.69  $\pm$ 0.19 & 1 & 0 \\
$f_{2d}$ & 2456.98  $\pm$ 0.13 &  & 0.056 & 2.0 & 1.25  $\pm$ 0.16 & 1 & $+1$ \\
$f_{2e}$ & 2458.233  $\pm$ 0.097 &  & 0.075 & 2.6 &    & 1 &  \\
$f_{3a}$ & 1167.818  $\pm$ 0.007 &  & 1.079 & 37.7 & 9.80  $\pm$ 0.15 & 1 & $-1$ \\
$f_{3b}$ & 1177.615  $\pm$ 0.151 & 849.2 & 0.247 & 6.2$\dagger$ & 8.78  $\pm$ 0.16 & 1 & 0 \\
$f_{3c}$ & 1186.399  $\pm$ 0.060 &  & 0.618 & 14.9$\dagger$ &    & 1 & $+1$ \\
$f_{4a}$ & 935.706  $\pm$ 0.042 &  & 0.174 & 6.1 & 9.45  $\pm$ 0.06 & 1 & $-1$ \\
$f_{4b}$ & 945.159  $\pm$ 0.042 & 1058.0 & 0.174 & 6.1 & 8.68  $\pm$ 0.06 & 1 & 0 \\
$f_{4c}$ & 953.844  $\pm$ 0.044 &  & 0.165 & 5.7 &    & 1 & $+1$ \\
$f_{5a}$ & 870.468  $\pm$ 0.002 &  & 4.665 & 162.9 & 9.34  $\pm$ 0.05 & 1 & $-1$ \\
$f_{5b}$ & 879.803  $\pm$ 0.048 & 1136.6 & 0.153 & 5.4 & 8.66  $\pm$ 0.08 & 1 & 0 \\
$f_{5c}$ & 888.463  $\pm$ 0.066 &  & 0.111 & 3.9 &    & 1 & $+1$ \\
$f_{6a}$ & 1088.579  $\pm$ 0.005 &  & 1.394 & 48.7 & 13.26  $\pm$ 0.04 & 2 & ? \\
$f_{6b}$ & 1101.844  $\pm$ 0.036 & 907.6 & 0.203 & 7.1 &    & 2 & ? \\
$f_{7}$ & 835.797  $\pm$ 0.006 & 1196.5 & 1.278 & 44.6 &    &  &  \\
$f_{8}$ & 971.305  $\pm$ 0.010 & 1029.5 & 0.726 & 25.3 &    &  &  \\
$f_{9}$ & 906.287  $\pm$ 0.014 & 1103.4 & 0.505 & 17.6 &    &  &  \\
$f_{10}$ & 652.340  $\pm$ 0.017 & 1532.9 & 0.424 & 14.8 &    &  &  \\
$f_{11a}$ & 1206.599  $\pm$ 0.046 &  & 0.158 & 5.5 & 13.89  $\pm$ 0.10 & 2 & ? \\
$f_{11b}$ & 1234.375  $\pm$ 0.020 & 810.1 & 0.370 & 12.9 &    & 2 & ? \\
$f_{12}$ & 769.535  $\pm$ 0.020 & 1299.5 & 0.360 & 12.6 &    &  &  \\
$f_{13}$ & 1043.372  $\pm$ 0.026 & 958.4 & 0.280 & 9.8 &    &  &  \\
$f_{14}$ & 1146.033  $\pm$ 0.028 & 872.6 & 0.262 & 9.1 &    &  &  \\
$f_{15}$ & 1009.076  $\pm$ 0.032 & 991.0 & 0.228 & 8.0 &    &  &  \\
$f_{16}$ & 744.375  $\pm$ 0.032 & 1343.4 & 0.228 & 8.0 &    &  &  \\
$f_{17}$ & 805.181  $\pm$ 0.041 & 1242.0 & 0.177 & 6.2 &    &  &  \\
$f_{18}$ & 717.425  $\pm$ 0.045 & 1393.9 & 0.161 & 5.6 &    &  &  \\
$f_{5}-f_{7}$ & 34.868  $\pm$ 0.018 &  & 0.396 & 13.8 &    & comb &  \\
$f_{5}-f_{4}$ & 65.431  $\pm$ 0.039 &  & 0.187 & 6.5 &    & comb &  \\
$f_{8}-f_{5}$ & 101.009  $\pm$ 0.026 &  & 0.278 & 9.7 &    & comb &  \\
$f_{6}-f_{5}$ & 218.102  $\pm$ 0.018 &  & 0.414 & 14.4 &    & comb &  \\
$f_{5}-f_{3}$ & 296.352  $\pm$ 0.037 &  & 0.197 & 6.9 &    & comb &  \\
$2f_{5}$ & 1741.942  $\pm$ 0.013 &  & 0.574 & 20.1 &    & comb &  \\
$f_{3}+f_{5}$ & 2038.346  $\pm$ 0.025 &  & 0.288 & 10.0 &    & comb &  \\ 
\hline
\end{tabular*}
\end{table}

\end{document}